# DARWIN

Science Across Disciplines

A Proposal for the
Cosmic Vision
2015-2025
ESA Plan

This page intentionally left blank

## Table of Contents

This page intentionally left blank

# Executive Summary

The discovery of extra-solar planets is one of the greatest achievements of modern astronomy. There are now more than two hundred such objects known, and the recent detection of planets with masses approximately 5 times that of Earth demonstrates that extra-solar planets of low mass exist. In addition to providing a wealth of scientific information on the formation and structure of planetary systems, these discoveries capture the interest of both scientists and the wider public with the profound prospect of the search for life in the Universe.

We propose an L-type mission, called *Darwin*, whose primary goal is the study of terrestrial extra-solar planets and the search for life on them. By its very nature, *Darwin* advances the first Grand Theme of ESA's Cosmic Vision. Accomplishing the mission objectives will require collaborative *science across disciplines* ranging from planet formation and atmospheres to chemistry and biology, and these disciplines will reap profound rewards from their contributions to the *Darwin* mission.

*Darwin* is designed to detect rocky planets similar to the Earth and perform spectroscopic analysis of them at mid-infrared wavelengths (6 to 20 μm), where the most advantageous contrast ratio between star and planet occurs. The spectroscopy will characterize the physical and chemical state of the planetary atmospheres and search for evidence of biological activity. The baseline mission lasts 5 years and consists of approximately 200 individual target stars. Among these, 25 to 50 planetary systems can be studied spectroscopically, searching for gases such as $CO_2$, $H_2O$, $CH_4$ and $O_3$.

Extracting the faint emission of an Earth-like planet from the overwhelming flux of its host star requires exquisite angular resolution and wide dynamic range. At mid-infrared wavelengths, resolving the Habitable Zone – where liquid water can exist – would require a 100-meter class telescope. Filled apertures of this size in space are presently impractical, and in fact, studies have identified interferometric combination of individual spacecraft telescopes as the best technique. These spacecraft are separated by the required hundred-meter scale baselines, and their light is combined with *nulling interferometry*, whereby the glare of the host star can be cancelled. Such a facility can also operate in *constructive imaging* mode, allowing sensitive, high spatial resolution observations of general astrophysics targets, particularly when there is a bright source in the field of view.

ESA and NASA have studied a number of interferometer architectures over the past decade, with the goal of identifying a design that provides excellent scientific performance while minimizing cost and technical risk. These efforts, which teamed academic experts with industrial partners, have resulted in a convergence and consensus on a single mission architecture consisting of a *non-coplanar X-array*, called *Emma*, using four collector spacecraft and a single beam combiner spacecraft. *Emma's* great strength is its simplicity and greater sky coverage compared to other configurations. For example, this architecture eliminates the need for deployable optical components, and the collector spacecraft carry only a single mirror each.

In parallel with these system studies, ESA and NASA have funded technological developments that have produced important progress on key *Darwin* technologies. For example, nulling breadboards currently operating in Europe and in the USA routinely achieve null depths of $10^{-5}$, very close to *Darwin's* requirement. ESA and NASA-funded programs have also made significant advances in Formation Flying technologies, both through simulations using test benches and software, as well as via development of actual flight hardware for the PRISMA mission. With continuing support, the key technologies for *Darwin* should be mature by 2010.

A cost estimate indicates that *Darwin* is within the realistic resource envelope of a cooperation between different space agencies. Specifically, *NASA* and *JAXA*, have indicated their interest in the mission and their willingness to participate in the study phase. *Darwin* within the Cosmic Vision Program provides a timely opportunity for European investigators to play a major role in a mission that will ignite *intense interest in both the research community and the wider public.*

# 1. Introduction

We propose an ambitious space mission to discover and characterize Earth-like planets and to search for evidence of life on them. The *Darwin* mission, undertaken within the Cosmic Vision programme of ESA with international collaboration, will address the fundamental questions of Humankind's origin and our place in the Universe.

Imaginative thoughts of worlds other than our own, perhaps inhabited by exotic creatures, have been an integral part of our history and culture. Giants of classic civilization, such as Democritos of Abdera (460-371 BC), Epicurus of Samos (341-270 BC) and the medieval philosopher and theologian Giordano Bruno (1548-1600 AD) imagined that we might not be alone in the Universe. These great thinkers were following an ancient philosophical and theological tradition, but their ideas, exciting as they might seem, were not based on observational or experimental evidence.

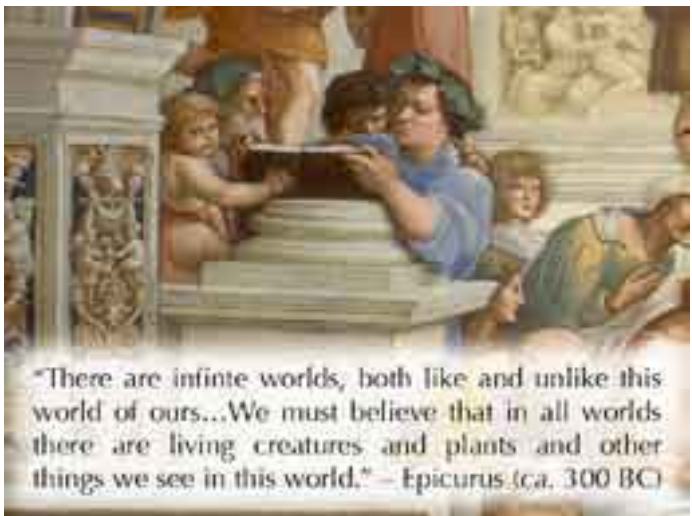

Figure 1.: The Greek philosopher Epicurus, in a detail from Raphael's "School of Athens" (1509)

Our understanding of our place in the Universe changed dramatically in 1995, when Michel Mayor and Didier Queloz of Geneva Observatory announced the discovery of an extra-solar planet around a star similar to our Sun. Geoff Marcy and Paul Butler in the United States soon confirmed their discovery, and the science of observational extrasolar planetology was born. The field has exploded in the last dozen years, resulting in over two hundred published planetary systems in 2007 (see http://exoplanet.eu/ and http://exoplanets.org/ for an up-to-date list).

Most of these systems contain one or more gas giant planets close, or very close, to their parent star, and thus, do not resemble our Solar System. Although very interesting in their own right, they do not directly address the possibility of other worlds like our own. Observational techniques continue to mature, however, and planets with sizes and mass similar to the Earth may soon be within reach. Almost monthly, reports appear on our television screens of "Super-Earths," several times more massive than our planet and potentially cloaked in life-supporting atmospheres. The most recent examples are GI 581c and GJ 436b.

Finding Earth analogues in terms of mass and size will be the focus of many ground and space-based research programmes in the coming decade. Finding evidence of habitability and life represents an even more exciting challenge. Semi-empirical estimates exist of the abundance of terrestrial planets, including the frequency of life and technologically advanced civilizations. Many of these assessments are based on Frank Drake's famous equation. Unfortunately, they are only educated guesses, not because the equation *per se* is incorrect, but rather because nearly all of its factors are essentially undetermined due to the lack of observational tests.

Thus, the basic questions remain open: "Are there planets like our Earth out there?" and "Do any of them contain life?" However, unlike our forebears going back more than 20 centuries, *we have the opportunity to address these questions using a truly scientific approach: observation.*

To characterize terrestrial exoplanets, we need to detect their light and analyse it by spectroscopy. There are two spectral domains where this can be done, the *visible* using the stellar light scattered by the planet, and the *thermal infrared (IR)* using the planetary emission. In Europe, we have chosen the latter for a variety of reasons, including the possibility to derive the planetary radius directly.

In order to extract the faint emission of a terrestrial planet from the overwhelming flux of its parent star, the planet must be spatially resolved. This requires extremely low background noise and an effective telescope diameter of ~100 m at thermal infrared wavelengths. An interferometer consisting of several free-flying spacecraft with baselines adjustable from 20 m to ~200 m is the best solution. Going to space offers several additional key advantages, including the ability to detect and characterize biologically important gases such as $H_2O$, $O_3$ and $CO_2$.

Building on the pioneering efforts of Bracewell (*Nat.* 274, 780; 1978) and Angel (Proc. of NGST conf., Baltimore, p.13-15; 1990), a team of European researchers proposed the *Darwin* concept to ESA in 1993 (Léger et al*.*, *Darwin* proposal to ESA Horizon 2000, 1993; *Icarus* 123, 249-255, 1996), and it has been studying it ever since. NASA has been advancing a similar concept, the Terrestrial Planet Finder Interferometer (TPF-I), since 1996 (Beichman et al., *JPL* Pub. 99-3; 1999). This proposal represents the cumulative effort and synthesis of these studies.

# 2. The Darwin Science Program

Searching for a phenomenon as subtle as life across parsecs of empty space may look hopeless at first glance, but considerable observational, laboratory, and theoretical effort over the past two decades is leading to the consensus that this is not the case. The *Darwin* science program is the logical climax of these efforts, and its goals are ambitious and profound: discovering other worlds like our own and examining them for evidence of extraterrestrial life.

We begin by asking what is life? A living being is a system that contains information and is able to replicate and evolve through random mutation and natural (Darwinian) selection (Brack, *Chem. Biodivers.* 4, 665-679, 2007). Although this definition appears overly generic (for example, it includes some computer viruses), consideration of possible storage media for life's information leads to a number of specific conclusions.

Macromolecules appear to be an excellent choice for information storage, replication, and evolution in a natural environment. Specifically, carbon chemistry is by far the richest and most flexible chemistry. The need for rapid reaction rates between macromolecules argues for a liquid solution medium. Based on physical and chemical properties as well as abundances in the universe, the most favourable, although not necessarily unique, path for life to take is then carbon chemistry in water solution (Owen, *ASSL* 83, 177, 1980). Fortunately, such chemistry produces a number of remotely detectable gaseous biological indicators in the planet's atmosphere.

The logic of the *Darwin* science program follows directly: we must search for habitable planets – those where liquid water can exist – and investigate their atmospheres for biosignatures, the gas products specific to the carbon macromolecule chemistry we call life.

## 2.1 Extrasolar Planetology in 2007

### 2.1.1 The Era of the Planet Hunters

The discovery of a planet orbiting the star 51 Pegasi (Mayor & Queloz, *Nat.* 378, p.355, 1995) marked the birth of a new field of astronomy: the study of extrasolar planetary systems around main sequence stars. Since then, more than 200 planets outside our own Solar System have been discovered. These planets most closely resemble the gas giant planets, with masses in the range 20 – 3 000 $M_\oplus$, but many of them are either in highly eccentric or very small (0.1 - 0.02 AU) orbits. The latter have surface temperatures up to 2 000 K, and are hence known as "Hot Jupiters".

The existence of Hot Jupiters can be explained by inward migration of planets formed at larger distances from their star, most likely due to tidal interactions with the circumstellar disk. We have also learned that Hot Jupiters preferentially form around higher metallicity stars: almost 15% of solar-type stars with metallicity greater than 1/3 that of the Sun possess at least one planet of Saturn mass or larger.

Despite considerable effort, no true solar system analogue has yet been found, and the lowest mass exoplanets range from 5 to 7 $M_\oplus$.

### 2.1.2 Planet Formation Theory

Planets form within disks of dust and gas orbiting newly born stars. Even though not all aspects are yet understood, growth from micrometer dust grains to planetary embryos through collisions is believed to be the key mechanism leading to the formation of at least terrestrial planets and possibly the cores of gas giants.

As these cores grow, they can become massive enough to gravitationally bind nebular gas. While this gas accretion proceeds slowly in the early phases, it eventually runs away when a critical mass is reached (~10 $M_\oplus$), allowing the formation of a gas giant within the lifetime of the gaseous disk. Terrestrial embryos, being closer to the star, have less material available and hence they empty their feeding zone before growing massive. They must then rely on distant gravitational perturbations to induce further collisions. As a result, the growth of terrestrial planets occurs on longer timescales than for the giants.

Caught by surprise at the time of their discovery, theory has since made considerable progress in understanding the known exoplanets. Extended core-accretion models can now be used to compute synthetic planet populations, allowing a statistical comparison with observations (see Figure 2.1). While these models are not specific to terrestrial planets (they are initialized with a seed of 0.6 $M_\oplus$), they demonstrate that if planetary embryos can form, only a small fraction of them will grow fast enough and big enough to eventually become giant planets. Given that we detect gas giants orbiting about 7% of the stars surveyed, *Darwin*'s harvest of terrestrial planet should be very significant indeed.

### 2.1.3 Habitability

The circumstellar *Habitable Zone (HZ) is defined as the region around a star within which an Earth-like planet can sustain liquid water on its surface*, a condition necessary for life. Within the HZ, starlight is sufficiently intense for a greenhouse atmosphere to maintain a surface temperature above 273 K, and low enough not to initiate runaway greenhouse con-ditions that can vaporize the whole water reservoir, allow photodissociation of water vapour and the loss of hydrogen to space (Kasting et al., *Icarus* 101, 108, 1993). The Continuous HZ is the region that remains habitable for durations longer than 1 billlion years. Figure 2.2 shows the limits of the Continuous HZ as a function of the stellar mass.

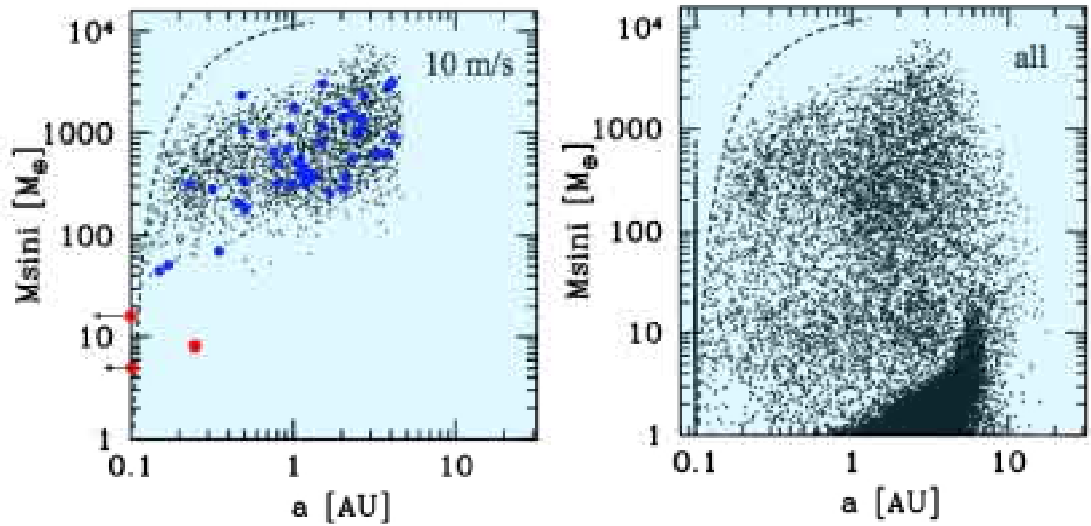

Figure 2.1: Extrasolar planet population synthesis. Models based on the core accretion scenario can be used to predict the expected planet population orbiting solar-type stars. Left, the planets predicted by the models and that are potentially detectable by Radial Velocity (10 m/s accuracy). The blue dots represent planets actually detected orbiting solar type stars, while the red dots are the 3 planets recently detected in orbit about the M dwarf Gl 581. Right, underlying population of planets. We note that the majority of embryos do not grow to become gas giants, leaving many detectable lower-mass planets.

Planets inside the HZ are not necessarily habitable. They can be too small, like Mars, to maintain active geology and to limit the gravitational escape of their atmospheres. They can be too massive, like HD69830d, which accreted a thick $H_2$-He envelope below which water cannot be liquid. However, planet formation models predict abundant *Earth-like* planets with the right range of masses (0.5 - 8 $M_\oplus$) and water abundances (0.01-10% by mass) (Raymond et al., *Icarus* 183, p.265 2006).

In order to determine if a planet in the HZ is *actually inhabited*, we need to search for biosignatures, spectral features that are specific to biological activity and which can be detected by remote sensing. An example is $O_2$ *producing photosynthesis*.

Chemolithotropic life, thriving in the interior of a planet without using stellar light, can still exist outside the HZ. The associated metabolisms – at least the ones we know on Earth – do not produce oxygen. However, chemolithotropic life relies on limited types of energy in the form of electron donors and acceptors. By comparison, the energy and electron donors for photosynthesis, sunlight and water respectively, are widely distributed, yield larger biological productivity, and can modify a whole planetary environment in a detectable way. A search for oxygenic photosynthesis would therefore yield a false negative on planets harbouring exclusively chemolithotropic life forms. Nevertheless, remote sensing of the spectroscopic signature of $O_2$ and its tracers (*e.g.* ozone) remains our best indicator of biological modification of a planet's atmosphere.

## 2.2 Characterizing Exoplanet Atmospheres and the Search for Life

The range of characteristics of planets found in the HZ of their star is likely to greatly exceed our experience with the objects in our Solar System. In order to study the habitability of the planets detected by *Darwin*, and to ascertain the biological origin of the measured atmospheric composition, we need a comprehensive picture of the observed planet and its atmosphere.

In addition to providing a more favourable planet-star contrast and some potential biosignatures, observations at mid-IR (MIR) wavelengths allow crucial chemical, physical and climatic diagnostics, even at moderate spectral resolution. For example, the infrared light curve, that is, the variation of the integrated thermal emission with location on the orbit, reveals whether the detected planet possesses a dense atmosphere, which reduces the day-night temperature difference and may provide conditions suitable for life.

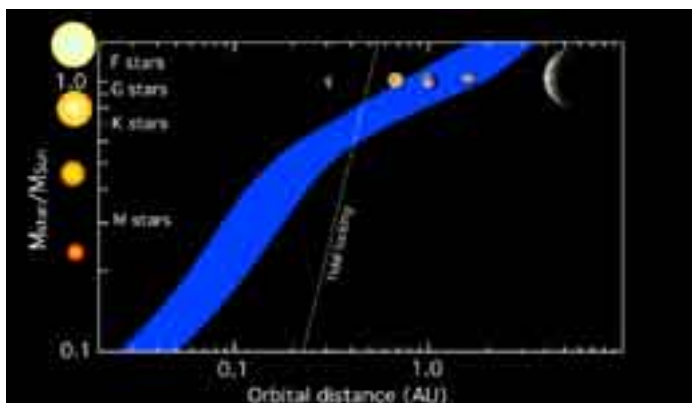

Figure 2.2: Continuous Habitable Zone (blue region) around M, K, G, and F stars. The region around the Sun that remains habitable during at least 5 Gyrs extends from approximately 0.76 to 1.63 AU.

A low-resolution spectrum spanning the 6-20 µm region allows us to measure the effective temperature $T_{eff}$ of the planet, and thus its radius $R_{pl}$ and albedo. Low-resolution mid-IR observations will also reveal the effects of greenhouse gases, including $CO_2$ and $H_2O$.

Within the HZ, the partial pressure of $CO_2$ and $H_2O$ at the surface of an Earth-like planet is a function of the distance from the star. Water vapour is a major constituent of the atmosphere for planets between 0.84 AU (inner edge of the HZ) and 0.95 AU. Figure 2.3 shows the estimated evolution of the $H_2O$, $O_3$ and $CO_2$ features in the spectra of an Earth-like planet as a function of its location in the HZ. Carbon dioxide is a tracer for the inner region of the HZ and becomes an abundant gas further out.

Planets such as Venus, closer to their star than the HZ, can lose their water reservoir and accumulate a thick $CO_2$ atmosphere. Such planets can be identified as uninhabitable by the absence of water and by the high-pressure $CO_2$ absorption bands between 9 and 11 µm.

*Darwin* will test the theory of habitability versus orbital distance by correlating the planets' spectral signature with orbital distance and comparing the results with grids of theoretical spectra, such as those shown in Figure 2.3.

2.2.1 Biosignatures

*Darwin* will have the ability to search for spectral signatures of life on planets found in the Habitable Zone of their star. Figure 2.4 shows that the mid-IR spectrum of Earth displays the 9.6 µm $O_3$ band, the 15 µm $CO_2$ band, the 6.3 µm $H_2O$ band and the $H_2O$ rotational band that extends longward of 12 µm. The Earth's spectrum is clearly distinct from that of Mars and Venus, which display the $CO_2$ feature only.

The combined appearance of the $O_3$, $H_2O$, and $CO_2$ absorption bands is the most robust and well-stud-ied signature of biological activity (Schindler and Kasting, *ESA SP* 451, 159, 2000; Selsis et al., *A&A* 388, 985, 2002; Des Marais et al., *Astrobio.* 2, 153, 2002). Despite variations in line shape and depth, atmospheric models demonstrate that these bands could be readily detected with a spectral resolution of 10–25 in Earth analogues covering a broad range of ages and stellar hosts (Selsis, ESA SP 451, p.133, 2000; Segura et al., *Astrobio* 3, 689, 2003; Kaltenegger et al., *ApJ* 658, 598, 2007).

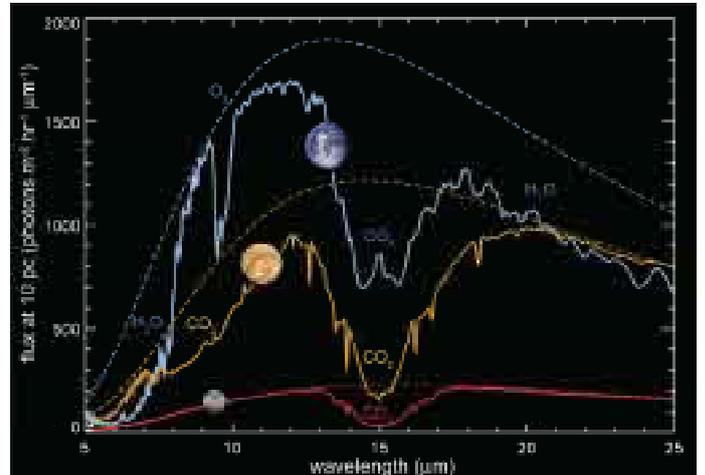

Figure 2.4: Mid-IR spectra of Venus, the Earth and Mars as seen from 10 pc

The ozone absorption band is observable for $O_2$ concentrations higher than 0.1% of the present terrestrial atmospheric level (Segura et al., *Astrobio* 3, 689, 2003). The Earth's spectrum has displayed this feature for the past 2.5 billion years.

Other spectral features of potential biological interest include abundant methane ($CH_4$ at 7.4 µm), and species released as a consequence of biological fixation of nitrogen, such as ammonia ($NH_3$ at 6 and 9-11 µm), nitrous oxide ($N_2O$ at 7.8, 8.5 and 17 µm) and nitrogen dioxide ($NO_2$ at 6.2 µm). The presence of these compounds would be difficult to explain in the absence of biological processes. Methane and ammonia commonly appear in cold hydrogen-rich atmospheres, but they are not expected as abundant abiotic constituents of Earth-size planetary atmospheres at habitable orbital distances. Known abiotic routes do not produce nitrous oxide and nitrogen dioxide.

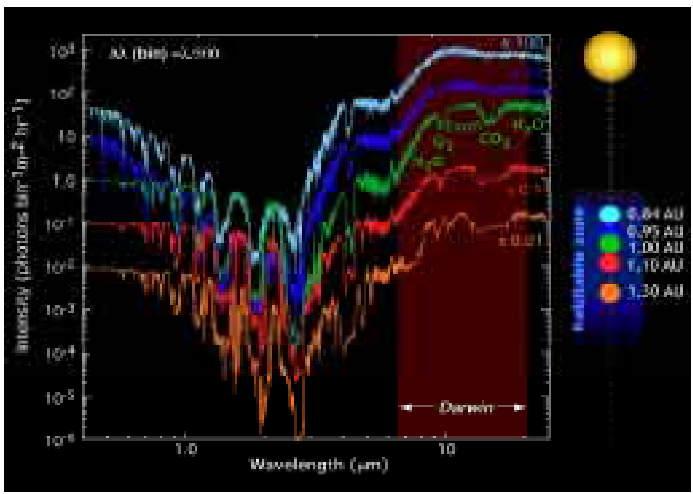

Figure 2.3: Synthetic spectra of an Earth-like planet computed at different orbital distance across the solar Habitable Zone (figure from Paillet et al., 2007, in press)

Methane, ammonia, nitrous oxide, and nitrogen dioxide do not produce measurable spectral signatures at low resolution for an exact Earth analogue. Nevertheless, they may reach observable concentrations in the atmosphere of exoplanets, due either to differences in the biosphere and the planetary environment, or because the planet is observed at a different evolutionary phase, as illustrated in Figure 2.5. Methane, for instance, was biologically maintained at observable concentrations during more than 2.7 billion years of Earth's history from about 3.5 until 0.8 Gyrs ago (Pavlov et al., *JGR* 105, 11981, 2000). During the 1.5 billion years following the rise of oxygen (2.4 Gyrs ago), the spectrum of the Earth featured deep methane absorption simultaneously with ozone. The detection of a reduced species, such as $CH_4$ or $NH_3$, together with $O_3$, is another robust indicator of biological activity (Lovelock, *Cosmic Search* 2, 2, 1980; Sagan et al., *Nat.* 365, 715, 1993).

The presence of $H_2O$, together with reduced species such as $CH_4$ or $NH_3$, would also be indicative of possible biological origin. Although a purely abiotic scenario could produce this mix of gases, such a planet would represent an important astrobiological target for future study. The presence of nitrous oxide ($N_2O$) and, more generally, any composition that cannot be reproduced by a self-consistent abiotic atmosphere model would merit follow-up.

Finally, if biology is involved in the geochemical cycles controlling atmospheric composition, as on Earth, greenhouse gases will likely be affected and sustained at a level compatible with a habitable climate. Whatever the nature of these greenhouse gases, *Darwin* will be able to see their effect by analyzing the planet's thermal emission. This is a powerful way to give the instrument the ability to characterize unexpected worlds.

## 2.3 Comparative (Exo) Planetology

Over the decade since the discovery of 51 Peg, we have grown to understand that planetary systems can be much more diverse than originally expected. Our Solar System represents a single sample of planets, after all. It is also clear that the current group of extrasolar planets, although diverse, is incomplete: as observational techniques have improved, we have pushed the lower limit to the detected masses closer and closer to the terrestrial range. In the coming decade, this trend will continue, and our understanding of the diversity of lower mass planets will be critical to the understanding of the formation of terrestrial planets in general, and of the Earth in particular.

Growing the sample of terrestrial planets from the three in our solar system to a statistically significant sample will represent a quantum leap in knowledge. And, just as 51 Peg created the discipline of observational extrasolar planetology, this effort will engender a new type of science: comparative exoplanetology for both the giant and terrestrial planets. It will allow for the first time a comparison of the orbital, physical and chemical characteristics of full planetary systems with our solar system and model predictions. Finally, this sample will also help answer one of the key questions related to *Darwin* science: How frequently are planets, which are located in or near the HZ, *truly habitable*?

### 2.3.1 Determination of Planetary Masses

*Darwin* can determine the radius but not the mass of planets. Ground-based radial velocity measure-ments can provide this information, however. The estimated error in mass determination is a function of planet mass, stellar type, visual magnitude, etc. Achieving adequate mass accuracy will be possible with instruments such as HARPS on 8-meter class telescopes, for a fraction of the discovered planets. Large planets (2 – 5 $M_\oplus$) around bright and/or late type stars are the best candidates (Figure 2.6).

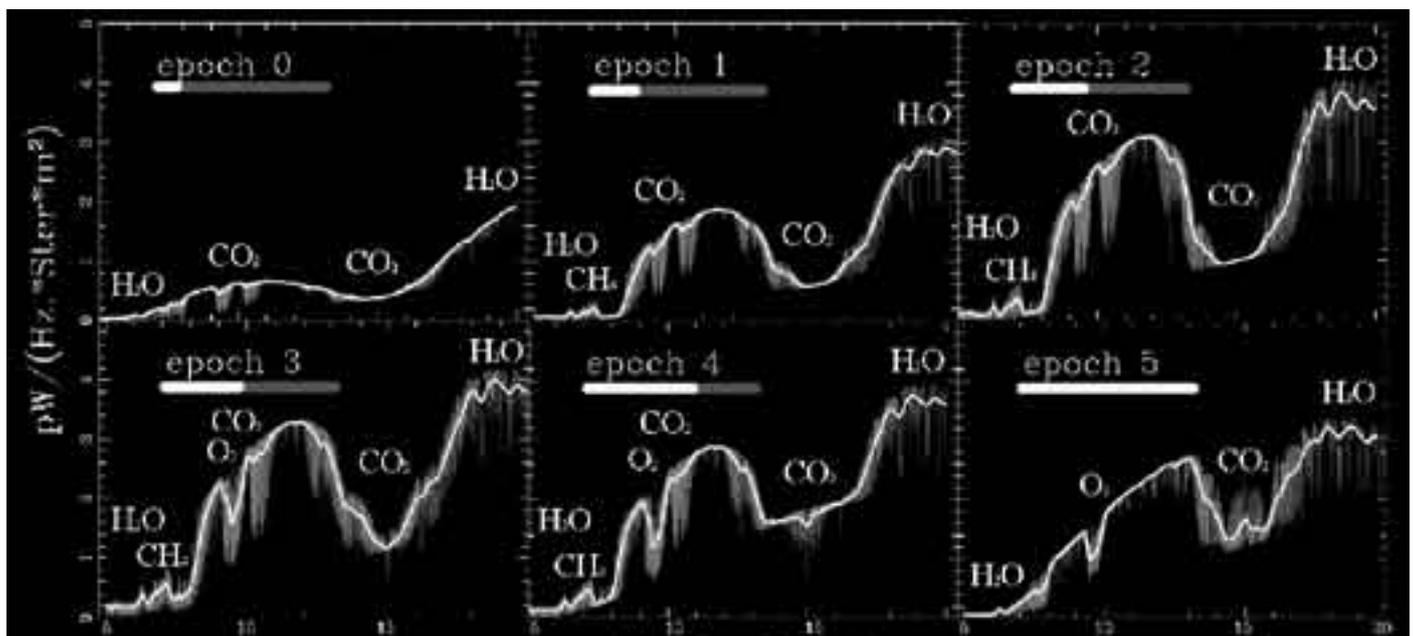

Figure 2.5: Mid-IR synthetic spectra of the Earth at six different stages of its evolution: 3.9, 3.0, 2.6, 2.0, 0.8 Gyrs ago and the present (figure from Kaltenegger et al. 2007, in press)

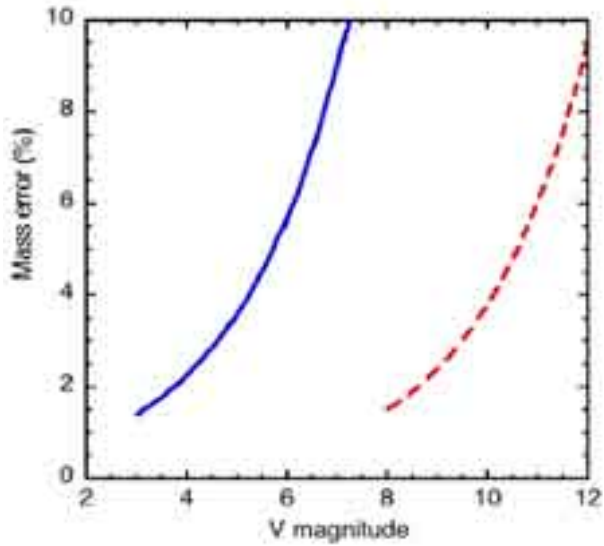

Figure 2.6: Estimates of the error on the mass determination of Darwin planets using the radial velocity technique for two cases: (Left) a 2 $M_\oplus$ planet around a G5V star, and (Right) a 5 $M_\oplus$ planet around a M3V star. Accurate determinations are possible for the brightest target stars within each spectral type. Assumed conditions are: quantum noise limited detection on an 8 m telescope, 40 h integration.

### 2.3.2 Habitability and the Water Reservoir

The origin and evolution of liquid water on the Earth is an ideal example of the type of puzzle that comparative exoplanetology will address. Our planet (thankfully) orbits in the Habitable Zone of our star, but at least some of the water on Earth must have been delivered by primordial icy planetesimal and/or water rich chondritic meteorites.

Did the early Earth capture these objects when they wandered into the inner solar system, or did our planet itself form further out and migrate inward? The answer is not clear at this point. What is clear is that habitability cannot just be reduced to a question of present-day location. The origin and fate of the water reservoir within the proto-planetary nebula is equally important.

By necessity, we have until now addressed this question using the very restricted sample of terrestrial planets in our own solar system: Venus, Earth and Mars. What have we learned? The in situ exploration of Mars and Venus taught us that all three planets probably evolved from relatively similar initial atmospheric conditions, most probably including a primordial liquid water reservoir. In all three cases, a thick $CO_2$ atmosphere and its associated greenhouse effect raised the surface temperature above the classical radiative equilibrium level associated with their distance to the Sun. This atmospheric greenhouse effect was critical for habitability on Earth at a time when the young Sun was approximately 30% fainter than it is today.

At some point in the past, the evolutionary paths of Venus, Earth, and Mars began to diverge. For Venus, the combination of the greenhouse effect and a progressively hotter Sun led to the vaporization of all liquid water into the atmosphere. After upward diffusion, the $H_2O$ was dissociated by UV radiation, causing the loss of hydrogen to space by gravitational escape and erosion. Venus is today a hot, dry, and uninhabitable planet.

In contrast, Mars apparently experienced a 500 million year episode with a warm, wet climate, before atmospheric loss and a steady decrease in surface temperature trapped the remaining water reservoir in the polar ice caps and subsurface permafrost. Thus, Mars also became uninhabitable, but retained a fraction of its water reservoir.

Earth apparently followed an intermediate and complex evolutionary path, which maintained its habitability for much of the past 4.6 billion years. Early on, a thick $CO_2$ atmosphere compensated for the young Sun's reduced luminosity, and as our star brightened, atmospheric $CO_2$ was progressively segregated into carbonate rocks by the combined action of the water cycle, erosion, sedimentation of carbonate deposits on the ocean floors, and partial recycling via plate tectonics. This feedback cycle, which appears unique in the Solar System, accounts for the preservation of Earth's oceans and habitability throughout most of its history.

Other external factors may influence a planet's atmosphere and its water reservoir. Specifically, we anticipate that the atmospheric evolution on planets orbiting lower mass M and K-type stars will be different. As illustrated in Figure 2.7, these planets experience denser stellar plasma environments (winds, Cosmic Mass Ejections). These stars also have longer active strong X-ray and Extreme UV periods compared to the Sun. Planets in the habitable zone of M and K stars can be partially or totally tidal-locked, which will produce different climatic conditions and weaker magnetic dynamos. Such dynamos protect the atmosphere against erosion by the stellar plasma flow. These differences compared to G-star HZ planets raise interesting questions regarding atmospheric escape, plate tectonics, magnetic dynamo generation and the possibility of complex biospheres.

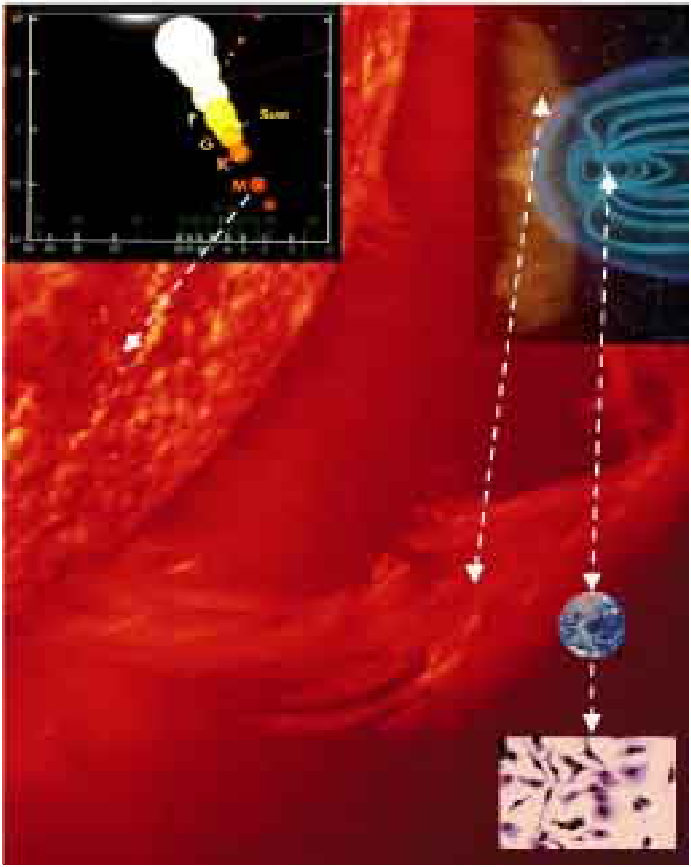

Figure 2.7: Mass ejections and winds from low mass M and K-type stars can erode the atmosphere of planets in the habitable zone. Darwin will study how such activity influences the magnetic dynamo, atmosphere, and biology of these planets.

The water reservoir on some habitable planets may be even more substantial than on Earth. For example, the recently proposed Ocean Planets, which consist of 50% silicates and 50% water by mass (Léger et al., *Icarus* 169, p.499, 2004; Selsis et al., *Icarus*, in press, 2007), could form further out than the water vapour condensation radius (~ 4 AU around a G star) and migrate to the HZ, or closer. Such objects would be a new class of planets, the terrestrial analogues of hot Jupiters and Neptunes. If Ocean Planets exist, *Darwin* will be able to characterize them in detail, for example determining their atmospheric composition.

### 2.3.3 Comparative Planetology with Darwin

With *Darwin*, the sample of terrestrial planets will be extended to our galactic neighbourhood, allowing us to study the relationship between habitability and three families of parameters:

- Stellar characteristics, including spectral type, metallicity, and if possible, age; our Solar System illustrates the importance of understanding the co-evolution of each candidate habitable planet and its star.

- Planetary system characteristics, particularly the distribution and the orbital characteristics of terrestrial and gas giant planets.

- The atmospheric composition of planets in the HZ. Here again, the Solar System sample points to the importance of ascertaining the relative abundance of the main volatile species: $CO_2$, $CH_4$, $H_2O$, $O_3$, $NH_3$, etc.

We can derive an approximate age of the planets based on the age of the host star. Asteroseismology measurements of the so-called "large" and "small" separations of the stellar oscillation eigenmodes allow a determination of the interval since the onset of hydrogen burning, and hence give the time since the star reached the Main Sequence. The accuracy is typically 5-10%, even for young stars. Space photometry missions can supply this valuable information.

The strategy for comparative exoplanetology will be as follows: First, a comparison of stellar characteristics with the nature of the planetary system will capture the diversity of planetary systems. Then, *Darwin*'s spectroscopic data will reveal the range of atmospheric compositions in the *Habitable Zone*, a range that will be related to the initial chemical conditions in the proto-planetary nebula and, if stellar ages are available, to the effects of atmospheric evolution.

Correlating the general characteristics of the planetary system with the atmospheres of the individual planets will illuminate the interplay between gas giants and terrestrial bodies and the role of migration. For example, recent numerical simulations predict that the scattering effect of giant planets on the population of planetesimals plays a key role in the collisional growth of terrestrial planets, their chemical composition and the build-up of their initial water reservoir.

Thus, *Darwin* will allow us to address the question of habitability from the complementary perspectives of the location of Earth-like planets with respect to their HZ, and of the origin, diversity and evolution of their water reservoirs.

## 2.4 High Angular Resolution Astronomy with Darwin

*Darwin's* long interferometric baselines and large collecting area make it a powerful instrument for general astrophysics. The mission has about the same sensitivity as JWST, and the angular resolution of VLTI in an instrument unencumbered by atmospheric opacity and thermal background.

The baseline instrumentation in *Darwin* will be able to observe general astrophysics targets whenever there is a bright point source, *e.g.* an unresolved star, in the field of view. This source is necessary to cophase the input pupils. Some science programs will profit from just a few visibility measurements, while others will require numerous observations and complete aperture synthesis image reconstruction. Section 4.4 gives further details.

Taking full advantage of interferometric imaging with *Darwin* – that is observing any source on the sky – will require specialized and potentially costly add-on instrumentation to allow the cophasing of the array. General astrophysics is not the primary science mission, but it will be important in the planned assessment studies to evaluate the additional cost and risk of adding such full-sky capability. A subsequent trade-off against scientific interest will drive the decision on implementing full-sky imagery, or not.

The following sections present the general astrophysics program, indicating in each case the scientific questions that *Darwin* can address and specifying the instrumental needs. Section 4 discusses the associated technical requirements.

**Star Formation**

Stars are the fundamental building blocks of the baryonic universe. They provide the stable environment needed for the formation of planetary systems and, possibly, for the evolution of life. *Darwin* will impact our understanding of star formation in fundamental ways, for instance the *Jet-Disk Connection*. Forming stars launch powerful jets and bipolar outflows along the circumstellar disk rotation axis. The mission could reveal the nature of the driving mechanism by spatially resolving the jet-launching region. Are jets formed by ordinary stellar winds, the magnetic X-points where stellar magnetospheres interact with the circumstellar disk, or are they launched by magnetic fields entrained or dynamo-amplified in the disk itself? *[Imaging with a bright point source]*

**Planet Formation**

Theory predicts that planets form in *circumstellar disks* over a period of $10^6$ to $10^8$ years. *Darwin* could provide a wealth of information about planetary systems at various stages of their evolution, revealing the origin of planetary systems such as our own, and thus helping to place our solar system into context. The mission will be unique in being able to spatially resolve structures below 1 AU in nearby star forming regions, allowing us to witness directly the formation of terrestrial planets in the thermal IR (Figure 2.8). *[Imaging with a bright point source]*

Additional planet formation science includes:

- *Disk formation and evolution.* *Darwin* will place constraints on the overall disk structure. For example, theory predicts that the vertical scale height is sensitive to the grain size distribution within the disk. The mission measurements will directly constrain grain growth, settling, and mixing processes in the planet-forming region. *[Imaging with a bright point source]*

- *Disk Gaps within the Inner Few AU* The mid-IR spectral energy distribution of protoplanetary and debris disks points to the existence of gaps. *Darwin* will determine if this clearing is due to the influence of already-formed giant planets or if they are the result of viscous evolution, photo-evaporation, and dust grain growth. *[Imaging with a bright point source]*

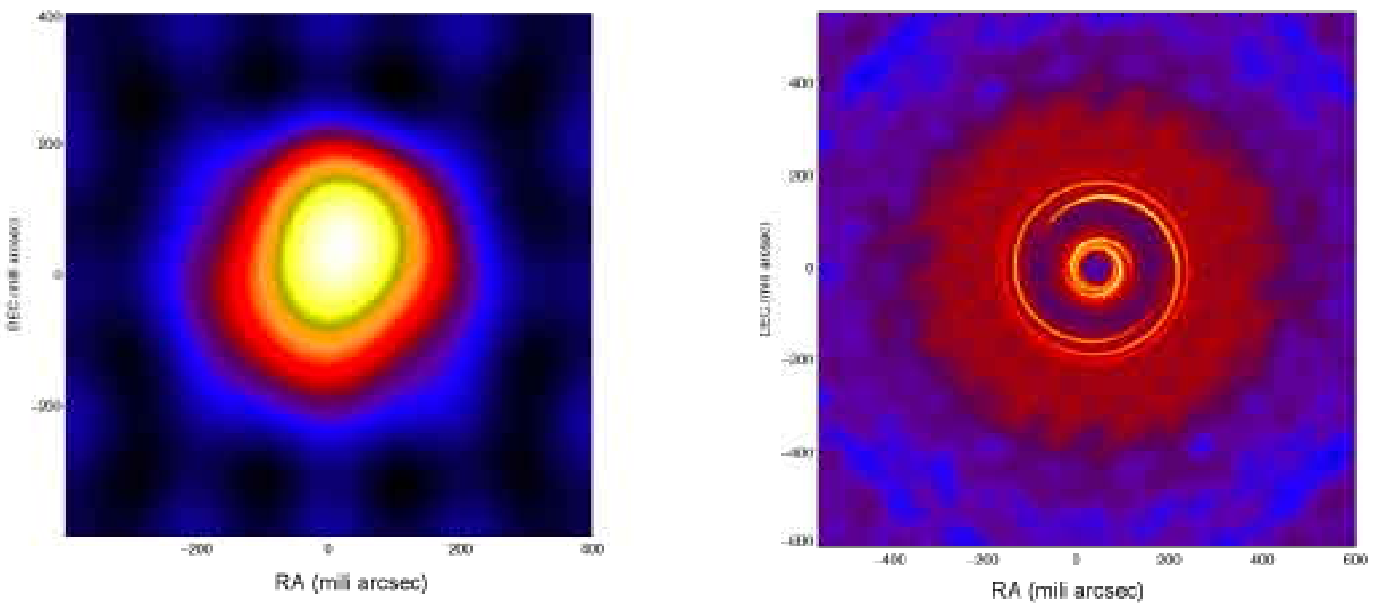

Figure 2.8: Simulation of a hot accretion disk in Taurus (140 pc) as seen by JWST (Left) and Darwin in its imaging mode as described in section 4.4 (Right). Simulated JWST and Darwin images are based on scaled models by D'Angelo et al. (2006) for the formation of a planet of one Jupiter mass at 5.2 AU, orbiting a solar type star. The most prominent features in the model are a gap along the planet's path and spiral wave patterns emanating from the Lagrangian points. Total observing time is 10 h. (courtesy Cor de Vries).

## Formation, Evolution, and Growth of Massive Black Holes

How do black holes (BH) form in galaxies? Do they form first, and trigger the birth of galaxies around them, or do galaxies form first and stimulate the formation of BHs? How do BHs grow? Do they grow via mergers as galaxies collide? Or do they accumulate their mass by hydrodynamic accretion from surrounding gas and stars in a single galaxy? *Darwin* could probe the *immediate environments of very different black holes*, ranging from very massive BH in different types of active galactic nuclei (AGN), to the massive black hole at the centre of our own Milky Way, down to BH associated with stellar remnants.

*Darwin* will make exquisite maps of the distribution of silicate dust, ices, and polycyclic aromatic hydrocarbons (PAHs) in weak *AGN* such as NGC 1068 out to a redshift of z=1-2. Brighter AGN can be mapped to a redshift of z=10, if they exist. For the first time, we will measure how the composition, heating, and dynamics of the dust disks change with redshift. This will provide a clear picture of when and how these tori and their associated massive BH grow during the epoch of galaxy formation. *[Imaging with a bright point source]*

*The Galactic centre:* The centre of our Galaxy contains the nearest massive black hole (3.6 x $10^6$ M☉), a uniquely dense star cluster containing more than $10^7$ stars pc$^{-3}$, and a remarkable group of high-mass stars with Wolf-Rayet-like properties. *Darwin* will be able to trace the distribution of lower mass stars and probe the distribution of dust and plasma in the immediate vicinity of the central BH. *[Imaging with no bright point source]*

## Galaxy Formation & Evolution

Galaxy evolution is a complicated process, in which gravity, hydrodynamics, and radiative heating and cooling all play a fundamental role. Measurements of the detailed spatial structure of very distant galaxies will place essential constraints on galaxy formation models. Both Darwin and JWST can observe these objects at 1-2 μm rest frame wavelength, the location of the peak of the spectral energy distribution of nearby galaxies. Unlike JWST, however, *Darwin* will resolve individual OB associations, massive star clusters, and their associated giant HII regions. By carefully selecting targets of a specific type, we can trace the evolution of galaxy structure as a function of redshift and environment. The evolution of metallicity with cosmic age and redshift can be mapped using various diagnostics, including molecular tracers, ices, PAH bands, and noble gas lines that are in the (6 – 20 μm) band. Figure 2.9 shows an example of the mapping power of the mission. *[Imaging with no bright source]*

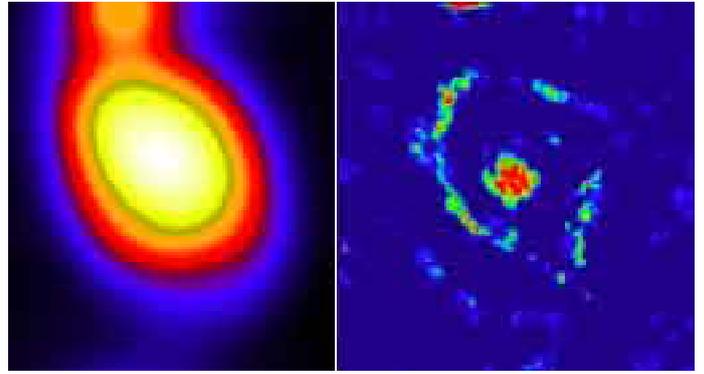

Figure 2.9: Simulated images of an M51-type galaxy at z=3, as observed with JWST (left) and Darwin (right).

## The First Generation of Stars

The first stars formed in the early universe are thought to be quite different from the stars present today. The absence of metals reduced the opacity, allowing this first generation of stars to accumulate more gas and hence be considerably more massive (100 to 1000 M☉) and hotter than their modern counterparts. The first stars must have had a dramatic impact on their environment, creating giant HII regions whose red-shifted hydrogen and helium emission lines should be readily observable by *Darwin*. While JWST is expected to make the first detections of galaxies containing these "Population III" stars, *Darwin* will resolve scales of order 10 to 100 pc at all redshifts, providing the hundred-fold gain needed to resolve these primordial HII regions. *Darwin* will also test the current paradigm for the formation of the first stars. Are they truly isolated, single objects that have inhibited the formation of other stars in their vicinity, or are they surrounded by young clusters of stars? *[Imaging with no bright source]*

## Other Important Science

The Darwin general astrophysics programme could include a number of additional key science targets:

- *Our home planetary system:* Darwin will easily measure the diameters and properties of Kuiper Belt Objects, moons, asteroids, and cometary nuclei. Low-resolution spectrophotometry will constrain the nature of their surfaces, atmospheres, and environments. *[A few visibility measurements are informative]*

- *AGB stars:* Darwin will provide detailed maps of the distribution of dust and gas within the envelopes of oxygen-rich (M-type) and carbon-rich (C-type) AGB stars, in environments as extreme as the Galactic centre. *[Imaging with no bright point source]*

- *Supernovae:* Darwin will image the formation and evolution of dust, atoms and ions in supernova ejecta, and trace the structure of

the circumstellar environment into which the blast is propagating. *[Imaging with a bright point source]*

- *Dark matter & dark energy:* Darwin studies of gravitational lensing by galaxy clusters, AGN, and ordinary galaxies will place unprecedented constraints on the structure of dark matter haloes at sub-kpc scales. *[Imaging with no bright point source]*

## 2.5 Synergies with other Disciplines

The primary *Darwin* science objective is inherently multi-disciplinary in character, uniting astrophysics with chemistry and biology. Often referred to as astrobiology, this interdisciplinary field also includes molecular biology, celestial mechanics and planetary science, including the physics and chemistry of planetary atmospheres and the characterization of exoplanetary surfaces. Finally, climatologists and ecologists will have the opportunity to study global influences on a statistical basis, with special emphasis on Venus type planets that have undergone a hot runaway.

On the technological front, the mission will drive development in such widely differing fields as material sciences, optical design, and spacecraft Formation Flying.

## 2.6 Darwin's Role in the Cosmic Vision Programme

*Darwin* fits extraordinarily well into the first Grand Theme of the Cosmic Vision 2015-2025 program: **What are the conditions for planet formation and the emergence of life?** Specifically, the mission is explicitly designed to explore sub-topic **1.2: From exo-planets to biomarkers,** *i.e. Searching for planets around stars other than the Sun, looking for biomarkers in their atmospheres, and imaging them.*

*Darwin* will also contribute significantly to theme **1.1: From Gas and dust to stars and planets**, *i.e. map the birth of stars and planets by peering into the highly obscured cocoons where they form.* In its imaging mode, operating at mid-IR wavelengths and with unprecedented spatial resolution, the mission could penetrate the dust obscuring the birthplaces of stars and planets and reveal the detailed physical processes driving star and planet formation.

In addition, *Darwin* science also addresses theme **2: How does the Solar System work?** Understanding the physics and dynamics of other planetary systems will certainly help to unravel the secrets of our own Solar System, including the issue of its long-term stability.

With its unprecedented high angular resolution capability, *Darwin* will advance Cosmic Vision theme **3: What are the fundamental physical laws of the Universe?** For example, determining the proper motion of stars in the gravitational field of the Galactic Centre with high accuracy will determine the properties of the black hole and of gravity itself.

## 3. The *Darwin* Mission Profile

### 3.1 Baseline Mission Scope

The *Darwin* mission consists of two phases, search and spectral characterization, whose relative duration can be adjusted to optimize scientific return. During the search phase of the mission (nominally 2 years), *Darwin* will examine nearby stars for evidence of terrestrial planets. An identified planet should be observed at least 3 times in order to characterize its orbit. The number of stars that can be searched depends on the level of zodiacal light in the system and the diameter of the collector telescopes. As a baseline, we estimate this number under the assumption of a mean exozodiacal density 3 times that in the Solar System and collecting diameters of 2 m. Over 200 stars can be screened under these conditions (section 4.3.3). The mission focuses on Solar type stars, including the F, G, K and some M spectral types.

The number of expected planetary detections depends upon the mean number of terrestrial planets per star in the habitable zone, $\eta_\oplus$. Our present understanding of terrestrial planet formation (section 2.1.2) and our Solar System, where there are 2 such planets (Earth & Mars) and one close to the HZ (Venus), point to a fairly high abundance of terrestrial planets. We assume hereafter that $\eta_\oplus = 1$. The COROT mission should reveal the abundance of small hot planets, and *Kepler* will evaluate $\eta_\oplus$ as well as the size distribution of these objects several years before *Darwin* flies. These inputs will allow refinement of *Darwin*'s observing strategy well in advance of launch.

During the characterization phase of the mission (nominally 3 years), *Darwin* will acquire spectra of each detected planet at a resolution of 20 and with sufficient signal-to-noise to measure the equivalent widths of $CO_2$, $H_2O$, and $O_3$ with a precision of 20% if they are in abundances similar to those in the Earth's atmosphere.

Spectroscopy is more time consuming than detection. With $\eta_\oplus = 1$, only a fraction of the detected planets can be studied spectroscopically. As shown in section 4.3.3 for Earth-sized planets, *Darwin* can perform spectroscopy of $CO_2$ and $O_3$ on about 50 planets and of $H_2O$ on about 25 planets during the nominal 3-year characterization phase. Note that the mission profile retains flexibility, and optimization of the spectroscopy phase will be possible based on early results from the detection phase.

The general astrophysics program, if adopted, will comprise 10% to 20% of the mission time. The primary science segment would then be reduced accordingly, with limited impact on its outcome.

## 3.2 Extended Mission Scope

An extension of the mission to 10 years will depend on the results gathered during the first 5 years. Such an extension could be valuable to observe more M stars, since only 10% of the baseline time is attributed to them. An extended mission would also permit a search for big planets around a significantly larger sample of stars, and additional measurements of the most interesting targets already studied.

## 4. Mission Design

### 4.1 The Darwin Concept and Its Evolution

In order to disentangle the faint emission of an Earth-like planet from the overwhelming flux of its host star, the planetary system needs to be spatially resolved. This, in turn, requires an instrument up to 100 metres in diameter when operating at mid-IR wavelengths, since the angular size of the *Habitable Zone*s around *Darwin* target stars ranges between 10 and 100 mas. A monolithic telescope of this size is presently not feasible, particularly since the observatory must be space-borne and cooled to provide continuous coverage and sensitivity between 6 and 20 μm.

As a result, interferometry has been identified as the best-suited technique to achieve mid-IR spectroscopy of Earth-like planets around nearby stars. In his pioneering paper, Bracewell (*Nat.* 274, 780, 1978) suggested that applying a π phase shift between the light collected by two telescopes could be used to cancel out the on-axis star, while allowing the signal from an off-axis planet to pass through (Figure 4.1). This technique, referred to as *nulling interferometry*, has been at the heart of the *Darwin* concept since its origin (Léger et al., *Darwin* proposal to ESA Horizon 2000 call, 1993) and many improvements have been studied since that date.

In addition to the planetary flux, a number of spurious sources contribute to the signal at the destructive output of the Bracewell interferometer:

- Residual star light, referred to as *stellar leakage*, caused by the finite size of the stellar photosphere and by imperfect efficiency of the interferometer;

- The *local zodiacal background*, produced by the disk of warm dust particles that surround our Sun and radiate at infrared wavelengths;

- The *exozodiacal light*, arising from the dust disk around the target star;

- The *instrumental background* produced by thermal emission within the instrument.

Bracewell's original suggestion of rotating the array of telescopes can help disentangle the various contributions. The planet signal would then be temporally modulated by alternately crossing high and low transmission regions, while the stellar signal and the background emission remain constant (except for the exozodiacal emission). Unfortunately, this level of modulation is not sufficient to achieve *Darwin's* goals, prompting a series of improvements to the strategy, including:

- Breaking the symmetry of the array to cancel all centro-symmetric sources, including the stellar leakage and the local and exozodiacal emission;
- Performing faster modulation of the planet signal via internal modulation between the outputs of sub-interferometers

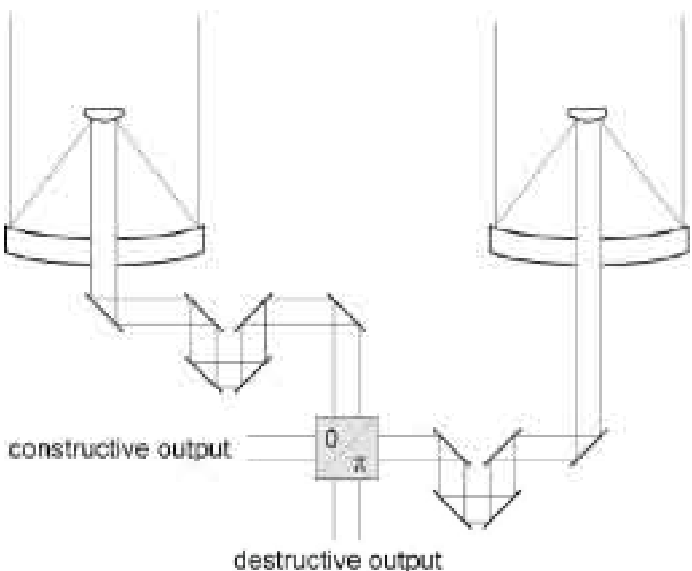
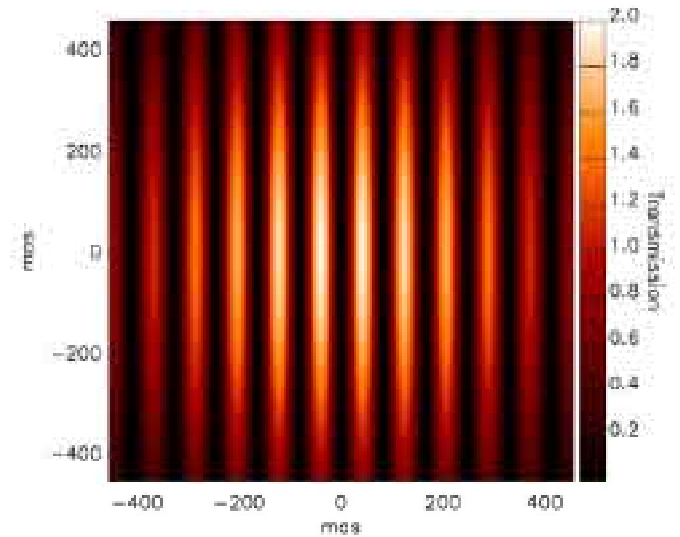

Figure 4.1: Left, principle of a two-telescope Bracewell nulling interferometer. Right, associated transmission map, displayed for λ=10 μm and a 25-m baseline array. This fringe pattern is effectively projected onto the sky, blocking some regions while transmitting others.

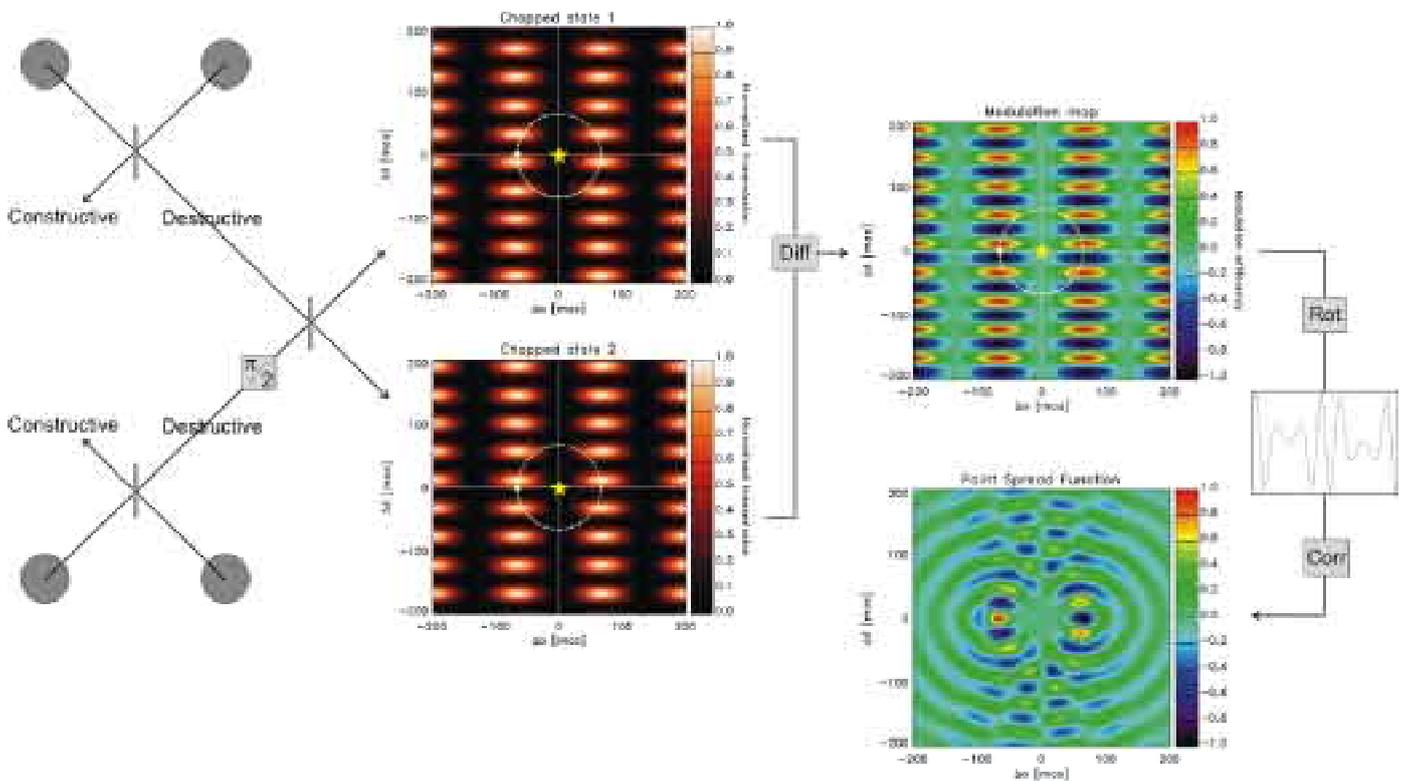

Figure 4.2: Phase chopping for the X-array, a four-element rectangular configuration of telescopes (see Section 4.2). Combining the beams with different phases produces two conjugated chopped states, which are used to extract the planetary signal from the background. Array rotation then locates the planet by cross-correlation of the modulated chopped signal with a template.

Merging of these two ideas has led to the concept of *phase chopping* (Mennesson et al., *Icarus* 178, 570, 2005, Woolf and Angel ASP Conf Series 119, p.285, 1997), which is now regarded as a mandatory feature in space-based nulling interferometry. Figure 4.2 illustrates the principle. The outputs of two Bracewell interferometers are combined with opposite phase shifts ($\pm\pi/2$) to produce two "chopped states," which are mirrored with respect to the optical axis. Taking the difference of the photon rates obtained in the two chopped states gives the chopped response of the array, represented by the modulation map. This chopping process removes all centro-symmetric sources, including the stellar leakage and the exozodiacal emission.

Because the modulation efficiency varies across the field-of-view, the planet can only be localised and characterised through an additional level of modulation, provided by array rotation with a typical period of one day. The variation of the chopped planet photon rate with the rotation angle of the array appears at the extreme right of Figure 4.2. These data must be inverted to obtain the fluxes and locations of any planets that are present. The most common approach is correlation mapping, which is closely related to the Fourier transform used for standard image synthesis. The result is a correlation map, displayed for a single point source in the lower right part of the figure. This represents the point spread function (PSF) of the array.

This process, illustrated here for a single wavelength, is repeated across the waveband, and the maps are co-added to obtain the net correlation map. The broad range of wavelengths planned for *Darwin* greatly extends the spatial frequency coverage of the array, suppressing the side lobes of the PSF (section 4.3.4).

A dozen array configurations using phase chopping have been proposed and studied at ESA and NASA during the past decade. In 2004, the two agencies agreed on common figures of merit to evaluate their performance. The most important criteria are the modulation efficiency of the beam combination scheme, the structure of the PSF and its associated ability to handle multiple planets, the overall complexity of beam routing and combination, and finally, the number of stars that can be surveyed during the mission lifetime (section 4.3.3). Among the many configurations studied, the X-array has been identified as the most promising for the *Darwin* mission.

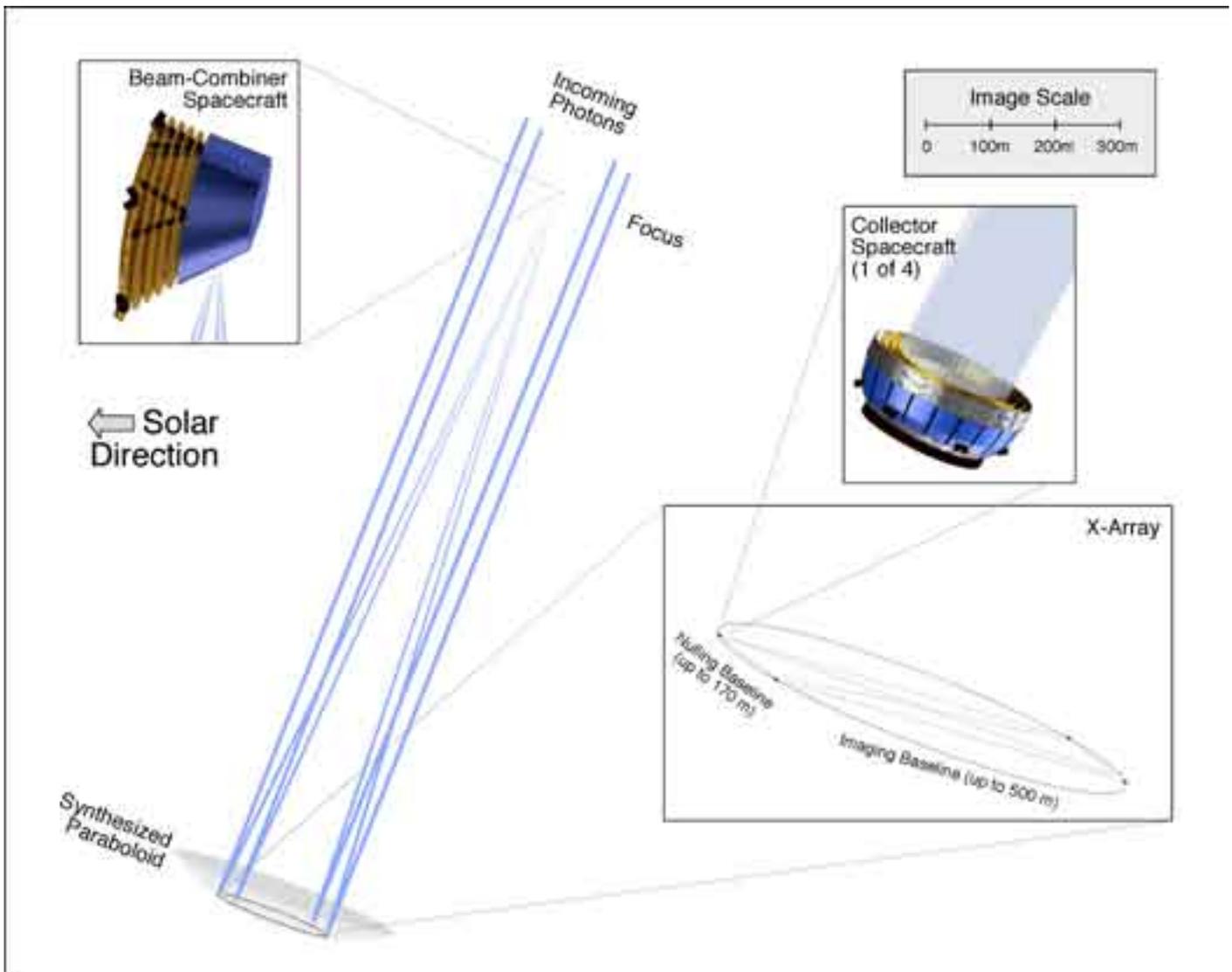

Figure 4.3: The Emma X-array configuration consists of 4 collector spacecraft and a beam combiner spacecraft. Spherical mirrors in the collectors form part of a large, synthetic paraboloid, feeding light to the beam combiner at its focus.

## 4.2 Mission Architecture

The desire for maximum mission efficiency, technical simplicity, and the ability to detect multiple planets around as many stars as possible has guided the selection of mission architecture. Additional top-level requirements include:

- Two observing modes: *nulling* for extrasolar planet detection and spectroscopy, and *constructive imaging* for general astrophysics;

- Placement at L2 for passive cooling and low ambient forces;

- Launch with a single Ariane 5 rocket or two Soyuz-ST/Fregat vehicles;

- The ability to search at least 250 candidate stars with an exozodiacal background of one zodi[1], or 150 stars with an exozodiacal background of 10 zodis;

- Detection and measurement of terrestrial atmosphere biosignatures as described in section 2.2 for at least 25 stars (with 1 zodi) or 15 stars (with 10 zodis);

- Time allocation of search as follows: G stars 50%, K stars 30%, F and M stars 10% each.

The effort to turn these requirements into a workable mission culminated in 2005-2006 with two parallel assessment studies of the *Darwin* mission. Two array architectures have been thoroughly investigated during these studies: the 4-telescope X-array and the 3-telescope TTN. These studies included the launch requirements, payload spacecraft, and the ground segment during which the actual mission science would be executed. Almost simultaneously, NASA/JPL initiated a similar study in the context of the Terrestrial Planet Finder Interferometer (TPF-I).

---

[1] A "zodi" is defined as the density of our local zodiacal dust disk and acts as a scaling factor for the integrated brightness of exozodiacal dust disks.

These efforts on both sides of the Atlantic have resulted in a convergence and consensus on mission architecture. The baseline for *Darwin* is a *non-coplanar,* or *Emma*[2]-type X-array, with four Collector Spacecraft (CS) and a single Beam Combiner Spacecraft (BCS). This process also identified a back-up option, in case unforeseen technical obstacles appear: a planar X-array.

### 4.2.1 The Emma X-Array Architecture

Figure 4.3 shows the non-coplanar Emma X-array. Four simple collector spacecraft fly in a rectangular formation and feed light to the beam combiner spacecraft located approximately 1200 m above the array. This arrangement allows baselines up to 170 m for nulling measurements and up to 500 m for the general astrophysics program.

The X-array configuration separates the nulling and imaging functions, thus allowing independent optimal tuning of the shorter dimension of the array for starlight suppression, and that of the longer dimension for resolving the planet. Most other configurations are partially degenerate for these functions. The X-array also lends itself naturally to techniques for removing *instability noise*, a key limit to the sensitivity of *Darwin* (section 4.3.2). The assessment studies settled on an imaging to nulling baseline ratio of 3:1, based on scientific and instrument design constraints. A somewhat larger ratio of 6:1 may improve performance by simplifying noise reduction in the post-processing of science images (section 4.3.2).

Each of the Collector Spacecraft (CS) contains a spherical mirror and no additional science-path optics (some additional components may be needed for configuration control). The four CS fly in formation to synthesize part of a larger paraboloid—the *Emma* configuration is a single, sparsely filled aperture. Flexing of the CS primary mirrors or deformable optics within the beam combiner spacecraft will conform the individual spheres to the larger paraboloid.

The Beam Combiner Spacecraft (BCS) flies near the focal point of this synthesized paraboloid. Beam combination takes place on a series of optical benches arranged within the BCS envelope. The necessary optical processing includes:

- Transfer optics and BCS/CS metrology;
- Correction and modulation, including optical delay lines, tip-tilt, deformable mirrors;
- Mirrors, wavefront sensors, and beam switching;

[2]Emma was the wife of Charles Darwin.

- Spectral separation, if necessary, to feed the science photons into 2 separate channels;
- Phase shifting, beam mixing ;
- Recombination, spectral dispersion and detection.

The collector and beam combiner spacecraft use sunshades for passive cooling to < 50 K. An additional refrigerator within the BCS cools the detector assembly to below 10 K.

Due to the configuration of the array and the need for solar avoidance, the instantaneous sky access is limited to an annulus with inner and outer half-angles of 46° and 83° centred on the anti-sun vector. This annulus transits the entire ecliptic circle during one year, giving access to almost the entire sky (Figure 4.4).

For launch, the collector and beam-combiner spacecraft are stacked within the fairing of an Ariane 5 ECA vehicle. The total mass (section 7) is significantly less than 6.6 tons, the launcher capability for delivery to L2. Table 4.1 lists key parameters of the *Darwin Emma* X-array. These values represent the results of the various assessment and system level studies conducted by ESA and NASA.

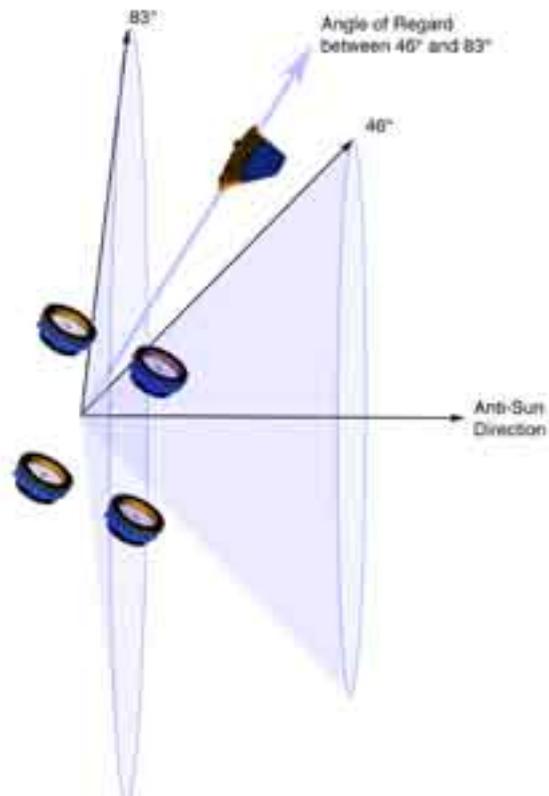

Figure 4.4: At any given time, the Emma X-Array can observe an annular region on the sky between 46° and 83° from the anti-Sun direction. During one Earth year, this annulus executes a complete circle, giving access to 99% of the celestial sphere.

Table 4.1: Key Darwin parameters

| Item | Value or Comment |
|---|---|
| Collector Spacecraft (CS) | 4 free-flyers, passively cooled to <50K |
| CS Optics | Lightweight spherical mirrors, diameter *ca.* 2.0 m, no deployables |
| CS Array Configuration | X-array with aspect ratio 3:1 – 6:1 (to be optimized) |
| Available Baselines | 7 m to 168 m Nulling, 20 m to 500 m Imaging option |
| Beam Combiner (BCS) | 1 free flying spacecraft, passively cooled to <50K |
| Beam Combiner Optics | Transfer, modulation, beam-mixing, recombination, spectroscopy |
| Detection | Mid-IR detector ca. 500 x 8 pixels for nulling, (300 x 300 for imaging option), cooled to < 10 K |
| Detector Cooling | Low vibration refrigerator, e.g. sorption coolers |
| Telemetry | Require ca. 1 GBit /s, direct downlink from BCS |
| Operating Wavelength | 6-20 µm. Includes $H_2O$, $O_3$, $CH_4$, $CO_2$ signatures |
| Field of View | Typically 1 arcsec at 10 µm |
| Null Depth | $10^{-5}$, stable over ~ 5 days |
| Angular Resolution | 5 milliarcsec at 10 µm for a 500 m baseline, scales inversely with wavelength |
| Spectral Resolution | 25 (possibly 300) for exo-planets; 300 for general astrophysics |
| Field of Regard | Annular region between 46° and 83° from anti-sun direction instantaneous, 99% of sky over one year |
| Target Stars | F, G, K , M, at least 150 (10 exo-zodis) or 220 (3 exo-zodis) |
| Mission Duration | 5 years baseline, extendable to 10 years |
| Mission Profile | Nominal 2 years detection, 3 years spectroscopy, flexible |
| Orbit | L2 halo orbit |
| Formation Flying | Radio Frequency and Laser controlled |
| Station Keeping | Field Effect Electric Propulsion (FEEP) or cold gas |
| Launch Vehicle | Single Ariane 5 ECA or 2 Soyuz-ST / Fregat |

## 4.3 Mission Performance

### 4.3.1 Detecting Earths

*Darwin*'s instruments will encounter a number of extraneous signals (see Figure 4.5), and the planetary flux must be extracted and analysed in the presence of these other components. This discrimination is performed by nulling the stellar light as much as possible, and by appropriate modulation (section 4.1) that produces a zero mean value for the different background sources. Unfortunately, modulation cannot eliminate the *quantum noise* (sometimes referred to as photon noise) associated with these sources.

For a given integration time, the signal is proportional to the number of planetary photons, and the quantum noise increases with the square root of the number of extraneous photons (Figure 4.5). As described in the next section, additional noise arises from imperfections in the system, such as stellar leakage. In order not to dominate the quantum noise, these imperfections must be very stable, with a flat Power Spectral Density, i.e. white noise, allowing the signal to noise ratio on the planet to increase as the square root of the integration time.

### 4.3.2 Instability Noise

To estimate mission performance in a realistic way, we must take into account the possible imperfections of the instrument. In the case of *Darwin*, the main imperfections result from vibrations and thermal drifts of the spacecraft, which in turn generate small fluctuations in the phase and amplitude of the input light beams. This produces instabilities in the interferometric null, a process similar to speckle noise in a coronagraph. The associated time-variable leakage of stellar photons, called *instability noise,* is generally not removed by phase chopping.

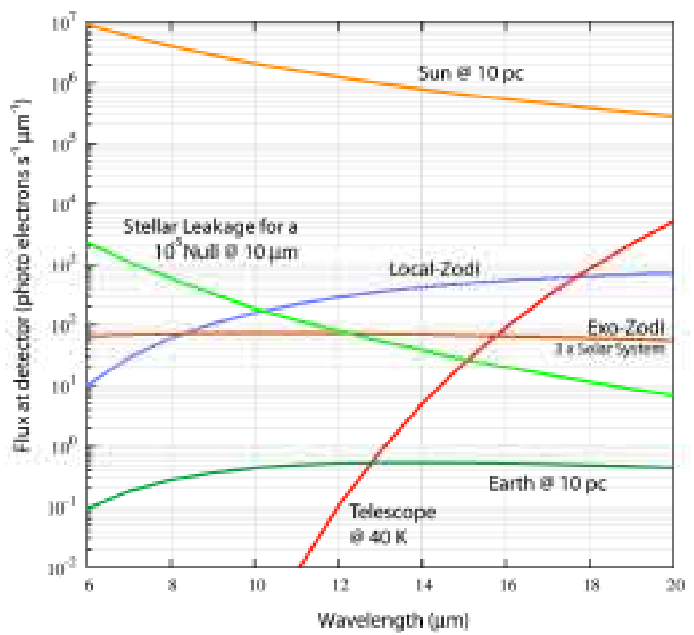

Figure 4.5: Different flux sources for an Earth analogue at 10 pc.

Reducing the contribution of instability noise to a harmless level, that is below other unavoidable sources, places strict requirements on configuration control: path length and amplitude variations should be less than 1.5 nm and 0.1%, respectively. This is only marginally compliant with state-of-the-art active control. Therefore, two techniques have been proposed and investigated to mitigate the influence of instability noise on mission performance.

The first solution, known as *spectral fitting*, is closely related to the particular arrangement of the X-array, where the nulling baselines are decoupled from the resolution baselines. Stretching the array while keeping the nulling baselines unchanged shrinks the fringe pattern considerably. Because the overall transmission pattern scales as the wavelength, the transmitted planet signal becomes a rapidly oscillating function of wavelength, an effect which can be used to disentangle it from the slowly varying instability noise pattern. The ability of the *Emma* architecture to adjust the aspect ratio of the X-array is valuable to implement this technique in an optimal way.

The second solution, known as *post-nulling calibration*, also relies heavily on the geometry of the X-array. The constructive outputs of the pair-wise nulling beam combiners, which contain mostly stellar light, can be used as reference beams to calibrate the final output of the interferometer. Only the stellar component of the output signal produces fringes when combined with these reference beams, and hence the stellar glare can be isolated.

These two mitigation techniques represent a decisive advantage of the X-array concept compared to other architectures.

### 4.3.3 Search Strategy and Performance

*Darwin* mission performance can be expressed in terms of the number of stars that can be screened for the presence of habitable planets, and the number of follow-up spectroscopic observations of planets that are possible.

The nominal mission is 5 years, with 2 years allocated to detection and 3 years for spectroscopy (possibly reduced by 10 to 20% if the imaging capability in constructive mode is implemented – see section 3.1). About 70% of mission time is spent collecting data, with the remainder dedicated to overheads, for example, moving the spacecraft to change the interferometer geometry. In order to secure an accurate identification in the search phase, we require that the probability for detecting an Earth-like planet in the HZ at a signal-to-noise ratio (SNR) of 5 be 90% or larger. We assume that the IR luminosity of planets in the HZ are identical to that of the Earth, independent of the stellar luminosity.

The *Darwin* target star catalogue was generated from the Hipparcos catalogue by examining the distance (< 25 pc), brightness (< 12 V-mag), spectral type (F, G, K, M main sequence stars), and multiplicity (no companion within 1 arcsec). The catalogue considers different interferometer architectures, since they have different sky access (see Section 4.2). The Emma design can observe 99% of the sky (section 4.2.1). The corresponding star catalogue contains 384 targets excluding M stars, and 625 stars total. Figure 4.6 shows some features of these stars.

ESA has conducted performance simulations for each star in the target catalogue, using the *Darwin*-Sim software to assess the integration time needed to reach the target signal to noise ratio (SNR) for detection and spectroscopy. These requirements are a SNR of 5 over the whole band for imaging in nulling mode, and a SNR of 10 from 7.2 to 20 μm for $H_2O$, $CO_2$ and $O_3$ spectroscopy[3], using a spectral resolution $\lambda/\Delta\lambda \geq 20$.

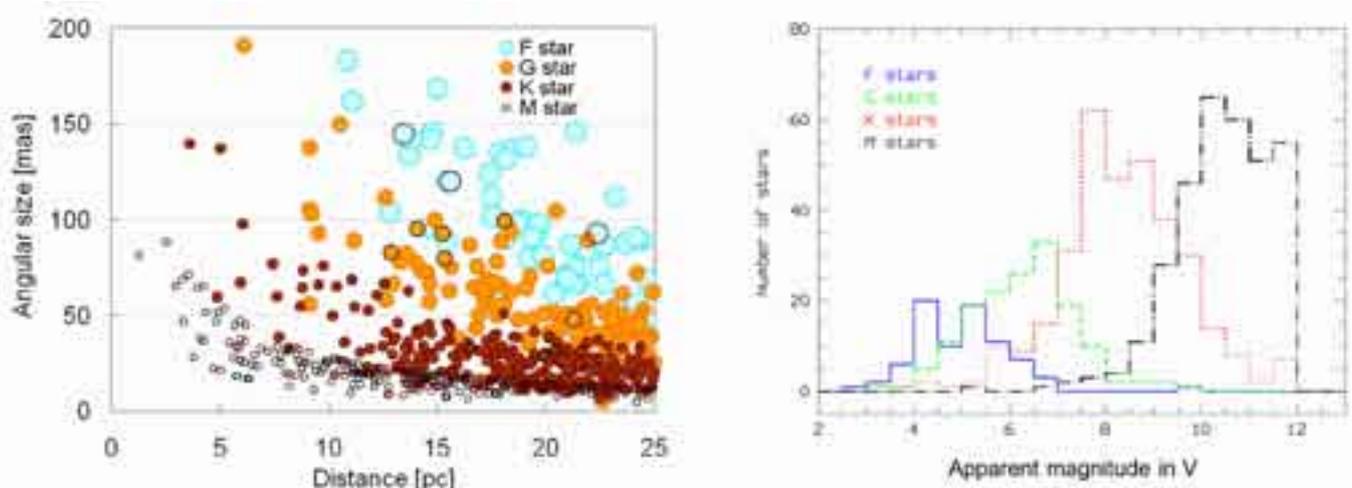

Figure 4.6: (left) Size of the Habitable Zone for the different spectral types of Darwin targets (Kaltenegger et al, 2007, in press). (right) Histogram of their apparent visual magnitude.

The level of exozodiacal emission is an important input parameter to these simulations. The amount of such emission around typical main sequence stars is largely unknown at present. There are successful ongoing programs using the Spitzer spacecraft and the Keck-I interferometer to evaluate stars with relatively bright zodiacal emission (*e.g.* young stars) and to relate this emission to stellar parameters. These measurements will allow pre-screening of *Darwin* targets to avoid objects with severe zodiacal emission. The LBT-I interferometer, and a possible future interferometer in Antarctica, will contribute to these efforts, which will conclude before Darwin's launch.

Under the assumption that the exozodiacal emission is symmetric around the target star, the chopping process will suppress it, and the exo-zodi will only contribute to the shot noise. The simulations presented below assume an exozodiacal density of 3 zodis.[4]

Using an Emma X-array (6:1 configuration) with 2-m diameter telescopes and assuming an optical throughput of 10% for the interferometer, we estimate that about 200 stars distributed among the four selected spectral types can be screened during the nominal 2-year survey (Table 4.2). *Darwin* will thus provide meaningful statistics on nearby planetary systems.

Figure 4.7 shows that nearby K and M dwarfs are the easiest targets in terms of Earth-like planet detection for a given integration time. This is because the thermal infrared luminosity of a planet in the Habitable Zone depends only on its size. On the other hand, the stellar luminosity is a strong function of its spectral type. This means that the star/planet contrast varies with spectral type. Compared to the case of the Sun and Earth, this contrast is two times higher for F stars, a factor of three lower for K stars, and more than an order of magnitude lower for M-dwarfs. Nevertheless, *Darwin* will focus on solar-type G type stars (50% of the observing time – see section 4.2), and a significant number of them can be screened and any discovered terrestrial planets studied (Table 4.2).

---
[3]The required SNR of 10 for water vapour detection is still under study. For $CO_2$ and $O_3$, a signal to noise ratio of 5 would actually be sufficient for a secure detection.
[4]In practice, exozodiacal densities below 10 times our local zodiacal cloud barely affect the overall shot noise level, while higher densities would significantly increase the required integration times.

Table 4.2: Number of stars and planets that can be studied

| Diameter | 1m | 2m | 3m |
|---|---|---|---|
| **Screened** | **76** | **218** | **405** |
| # F | 5 | 14 | 30 |
| # G | 15 | 53 | 100 |
| # K | 20 | 74 | 152 |
| # M | 36 | 77 | 123 |
| $CO_2$, $O_3$ | 17 | 49 | 87 |
| # F | 1 | 2 | 3 |
| # G | 4 | 8 | 15 |
| # K | 3 | 12 | 25 |
| # M | 9 | 27 | 44 |
| $H_2O$ | 14 | 24 | 43 |
| # F | 0 | 1 | 1 |
| # G | 2 | 4 | 7 |
| # K | 1 | 5 | 10 |
| # M | 11 | 14 | 25 |

Assuming that each nearby cool dwarf is surrounded by one rocky planet of one Earth radius within its *Habitable Zone*, we estimate that only a fraction of the detected planets can be fully characterised (i.e., examined for the presence of $H_2O$, $CO_2$ and $O_3$) during the subsequent 3-year spectroscopic phase. The numbers in Table 4.2 should be doubled or quadrupled for planets with radii 1.5 and 2 times that of the Earth, respectively. A comparable simulation effort at NASA using star count models confirms these predictions.

The diameter of the collector telescopes has an important influence on the overall mission performance. With 1-m telescopes, the number of targets screened would be reduced to about 75, while with 3-m mirrors, about two thirds of the star catalogue could be surveyed, at constant performance per object.

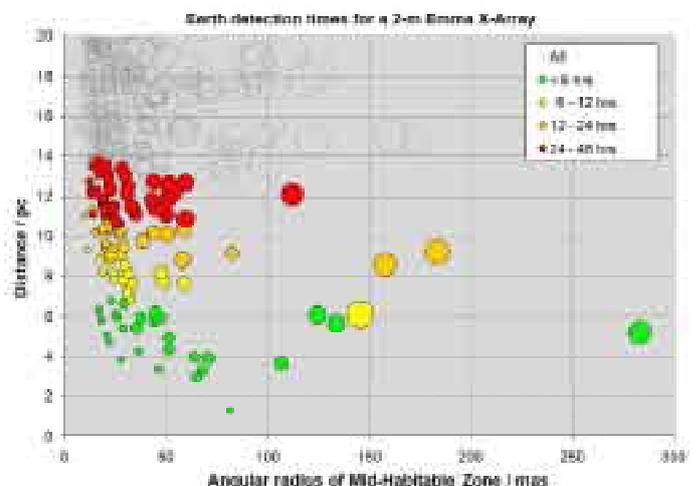

Figure 4.7: Integration time for a 5σ detection of Earth-sized planets around each of the Darwin candidate targets. The diameter of each symbol corresponds to the physical diameter of the star, and the colour relates to the integration time.

### 4.3.4 Image Reconstruction

The mission performance estimates in the preceding section are based solely on signal-to-noise ratio, disregarding the details of signal extraction and image reconstruction algorithms. In practice, data processing will be very important. For example, accurate orbit determination requires that the emission can be localised and tracked over time, while spectroscopy is only meaningful if the photons can be attributed to the right object. It is very important that *Darwin* can resolve the emission from the multiple sources that might be present in a stellar system, including planets, lumps in the exozodiacal dust emission, background objects, etc. Image fidelity depends on a high quality PSF, such as that shown for the X-Array in Figure 4.2. As mentioned before, this configuration has the major advantage of allowing a separation of the nulling and imaging baselines.

Software development for nulling imaging has been initiated on both sides of the Atlantic. In Europe, a Bayesian approach has been chosen. It takes advantage of all of the available information for image reconstruction, for example, incorporating positivity and smoothness of the spectra, a process called "regularization." Figure 4.8 illustrates how these constraints can enhance detection performance

## 4.4 Imaging for the General Astrophysics Program

Section 2.4 mentioned the importance of a bright source in the field to cophase the sub-pupils of the interferometer. The nature of the target and the science goal will determine the required instrumentation. We consider three different cases: visibility measurements with a bright source in the field of view, aperture synthesis imaging of targets with a bright source in the field, and aperture synthesis for targets with no bright source.

### 4.4.1 Visibility Measurements with a Bright Source in the Field of View

With minimal impact on the nulling recombiner, *Darwin* can carry out visibility ($V^2$) science with JWST-like sensitivity, as long as there is a $K \leq 13$ magnitude source in the field of view to stabilize the array. The modulus of the visibility provides simple size information about the target, for example, its radius assuming spherical morphology. The phase of the visibility gives shape information, such as deviations from spherical symmetry.

If the target spectrum is smooth, a few visibility measurements can be obtained rapidly, because *Darwin* can work simultaneously at several wavelengths (see below). The baseline recombiner could perform such measurements with very modest modification and minimum impact on mission cost. A capability for basic visibility measurement should therefore be implemented. Unfortunately, only a few types of targets can benefit from a limited number of $V^2$ observations (see section 2.4).

### 4.4.2 Aperture Synthesis for Targets with a Bright Source in the Field

To obtain a fully reconstructed image, the (u,v) plane must be filled by moving the array. A significant gain in efficiency can be realized if the spectrum of the target is smooth over the operating band of the instrument. The shorter wavelengths sample higher spatial frequencies, and the longer wavelengths lower spatial frequencies, all at the same array spacing. Figure 4.9 show an example of (u,v) coverage for circular trajectories of the individual spacecraft. A 100x100 image could be obtained with a few hundreds of positions, rather than the 10,000 positions normally required for a full spatial and spectral reconstruction. Table 4.3 lists the requirements for this mode.

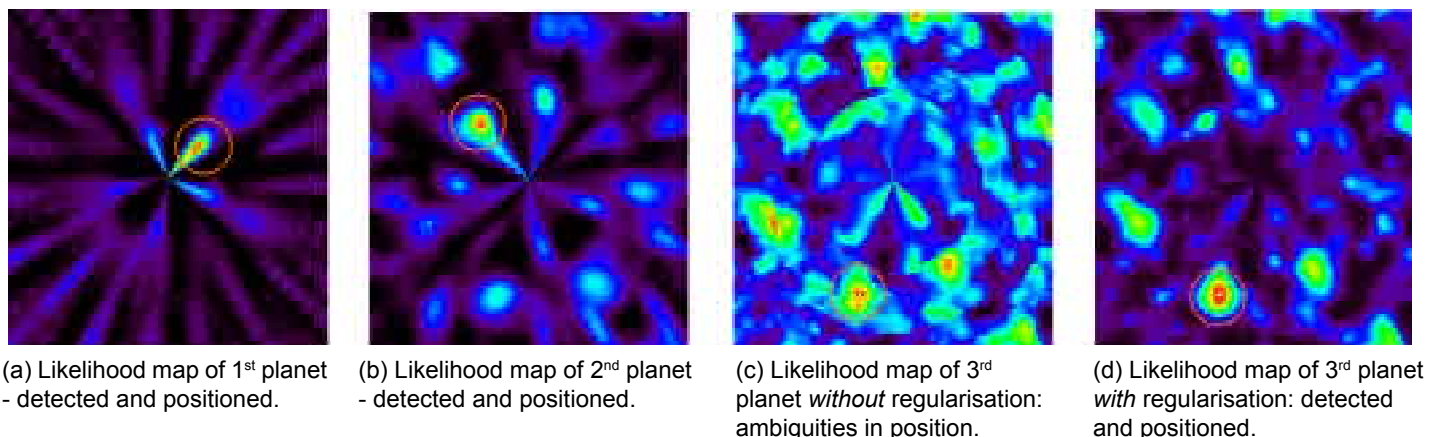

(a) Likelihood map of 1st planet - detected and positioned.

(b) Likelihood map of 2nd planet - detected and positioned.

(c) Likelihood map of 3rd planet *without* regularisation: ambiguities in position.

(d) Likelihood map of 3rd planet *with* regularisation: detected and positioned.

Figure 4.8: Likelihood maps of the successive detection of 3 planets located at 0.64, 1.1 and 1.8 AU from a star. Red indicates a higher probability, black a lower one. White is the highest. The spectral resolution is 15 and the SNR is 0.33 per spectral element. The 3rd and faintest planet is firmly detected and localised when regularisation is used.

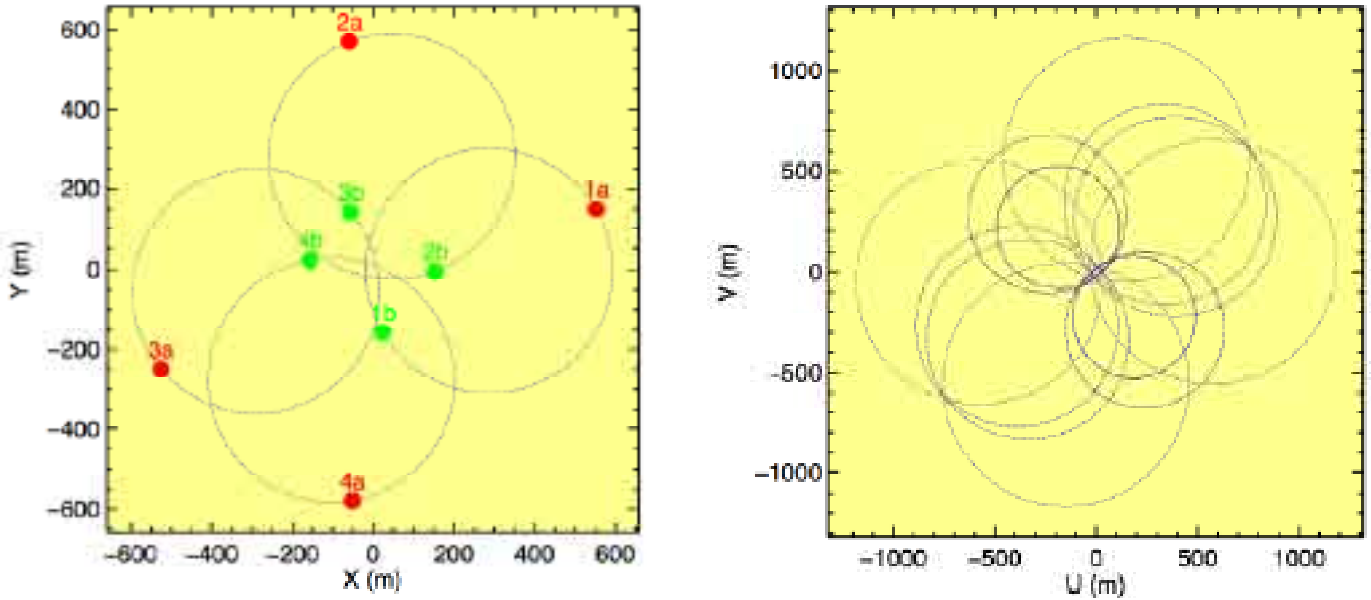

Figure 4.9: Possible satellite tracks and UV-coverage for 4 collectors. The satellites move independently on circular paths with the hub located at X=0, Y=0, continuously varying the dimensions of the array (left). The positions of the spacecraft are represented at 2 different epochs (a and b). The (u,v) plane is already well-filled after a single tour of the spacecraft (right).

*Sensitivity:* The 5$\sigma$, one hour, point source sensitivities for *Darwin* in 20% bandwidths centred at 8, 10, 13 and 17 µm are approximately 0.1, 0.25, 0.5 and 0.8 µJy, respectively. These sensitivities are comparable to those of JWST.

*Angular resolution* The maximum foreseen baselines are 500 metres, corresponding to a spatial resolution of 5 mas at 10 µm.

*Co-phasing:* With a stability time scale of 10 seconds for the array (Alcatel study, 2000), the sensitivity limit for self-fringe-tracking is about 10 mJy at 10 µm in a 0.5 arcsec aperture. This performance gives access to virtually all of the sources in the Spitzer SWIRE survey.

### 4.4.3 Aperture Synthesis for Targets with no Bright Source

For targets with no bright source in the field, the preferred option for co-phasing is the use of a nearby off-axis bright reference star (K $\leq$ 13). One way of doing this is to feed the K-band light of this star along with the 6-20 µm light of the target to the Beam Combiner Satellite, a so-called dual field configuration. Another option is to make the interferometer optically rigid using Kilometric Optical Gyros. These devices can maintain the phasing of the array between pointings at a reference star and the target field.

Clearly, this additional instrumentation may be much more demanding and expensive. A decision whether or not to add this capability will depend on an analysis undertaken during the study phase.

## 5. Science Operations and Archiving

### 5.1 Data Science Operations Architecture and Share of Responsibilities

The Science Operations Center (SOC) will be responsible for science mission planning, data processing, and data product distribution to the *Darwin* science team and the wider scientific community. Because the data acquisition and calibration requirements are very different for the planetary (nulling) and general astrophysics (imaging) missions, options for the SOC beyond ESOC need to be considered. Computer networking and remote presence through videoconferencing will play a central role, allowing responsibilities to be spread among a variety of network-nodes at several institutes throughout Europe.

### 5.2 Archive approach

The site for active and legacy archives is still to be determined. The archive should include values of the nulling transmission, visibilities, and reduced image data, including the accompanying calibration files. A quick-look facility will allow rapid assessment and review of the data. Compared to other contemporary missions, *Darwin*'s data volume will be relatively modest and should present no storage challenges.

Table 4.3: Aperture Synthesis imaging requirements for Darwin

| Item | Requirement | Goal |
|---|---|---|
| Maximum baseline [m] | 300 | 500 |
| Minimum baseline [m] | 20 | 10 |
| Field of view [resolution elements] | $100^2$ | $300^2$ |
| Dynamic range | 1 : 100 | 1 : 1000 |
| Spectral range [µm] | 6 - 20 | 4 - 30 |
| Spectral resolution | 300 | 3000 |

## 5.3 Proprietary Data Policy

Although the detailed rules of data access are still to be determined, we anticipate that there may be different policies for the primary science and general astrophysics programs. Specifically, the baseline mission (nulling interferometry) will be conducted by ESA in cooperation with a dedicated team of *Darwin* scientists. Data rights would then follow guidelines adopted by ESA for missions similar in character (e.g., GAIA). In general, the science team is obliged to reduce the data and make the results public within a stipulated time. A peer-review process will almost certainly determine the general astrophysics targets. Following a call for Open Time observations, ESA will accept proposals from a Lead Scientist, who will act as the contact point between the Agency and the proposing community. In this case, the commonly adopted proprietary period is one year from the time of data release.

# 6. Technology and Mission Roadmap for *Darwin*

## 6.1 Darwin's Technology Roadmap

### 6.1.1 Essential Technology Developments for Darwin

The pre-assessment study of *Darwin* by Alcatel in 2000, and the assessment study by TAS and Astrium in 2006, determined that there are no technology show stoppers for this ambitious mission. However, two key areas were identified that require focused attention and resources:

- *Formation Flying* of several spacecraft with relative position control of a few centimetres.
- The feasibility of *nulling interferometry* in the 6 - 20 μm range.

Based on the expected star/planet contrast ($1.5 \times 10^{-7}$ at 10 μm and $10^{-6}$ at 18 μm for a Sun-Earth analogue) and on evaluations of instability noise (section 4.3.2), the common conclusion of the industrial studies is that the null depth must be $10^{-5}$ on average, and that it must be sufficiently stable on timescales of days to allow the SNR to increase as the square root of time. This stability requirement translates into tight instrument control specifications. The two instability noise mitigation techniques presented in section 4.3.2 will allow a relaxation of these specifications, however. A thorough evaluation of the these methods and of the resulting instrumental stability requirements will be a key component of the technology development programme.

### 6.1.2 Current Status of Technology Development

Europe has devoted considerable resources, both intellectual and financial, to these technological issues since the initial Alcatel study. ESA has invested approximately 20 M€ since 2000, with a significant ramp-up in the last 2 years. Several tens of Technology Research Programs (TRPs) have been issued. NASA has run a parallel program in the USA. Most of the key technologies have been addressed and significant progress achieved.

In the area of Formation Flying (FF), the TRPs "Interferometer Constellation Control" (ICC1 and ICC2) have developed nonlinear, high fidelity navigation simulators. Algorithms for Interferometer Constellation Deployment at L2 have also been demonstrated. In the USA, analogous simulations and a 2D robotic breadboard (Figure 6.1) have shown the feasibility of formation flying. Finally, with the PRISMA mission being prepared for launch next year (section 6.2), Formation Flying is approaching Technology Readiness Level 5/6 (TRL 5/6).

The investment in nulling interferometry research over the past 7 years has brought the technology to TRL 4. The flight requirement is a null depth of $10^{-5}$ in the 6 – 20 μm domain. In Europe and at JPL, monochromatic experiments using IR lasers at 3.4 μm and 10.6 μm have yielded nulls equal to or significantly better than $10^{-5}$ (Figure 6.2). Broadband experiments have achieved nulls of $1.2 \times 10^{-5}$ for 32% bandwidth at 10 μm, closely approaching the flight requirement (Peters, 2007, http://planetquest.jpl.nasa.gov/TPF-I/). Clearly, the technology of nulling interferometry is nearing maturity, although it has not yet been demonstrated over the full *Darwin* bandwidth with the required depth and stability. Nevertheless, these results give us confidence that the mission

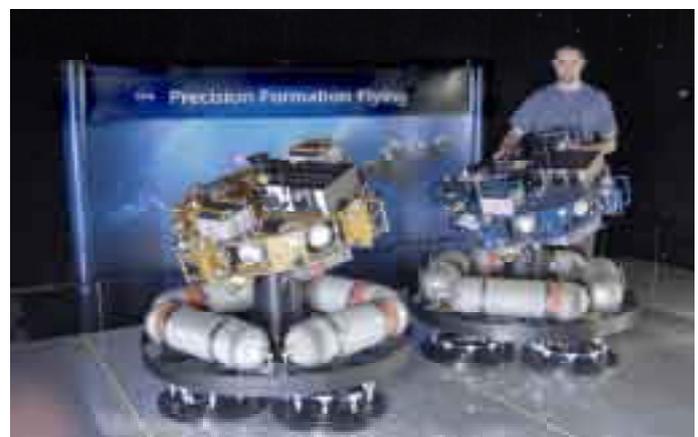

Figure 6.1: The 2 robots of the Formation Control Testbed at JPL. Each robot carries a mobile instrument platform (shown tilted), as well as canisters of compressed air to float the robot on a polished metal floor. The testbed has completed its functional testing and should achieve operational testing in 2007.

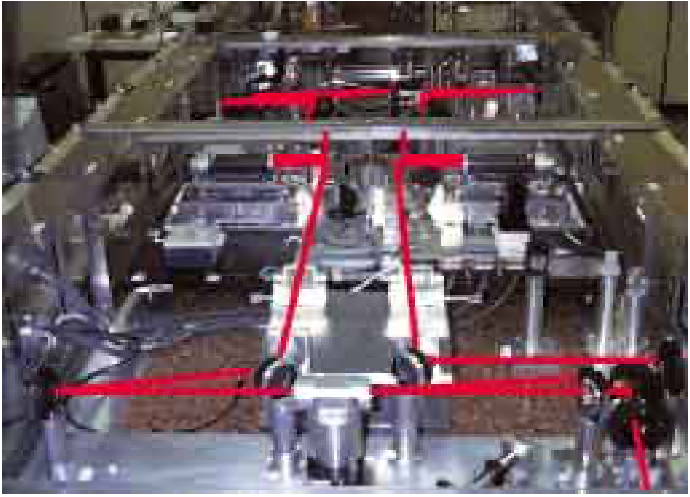 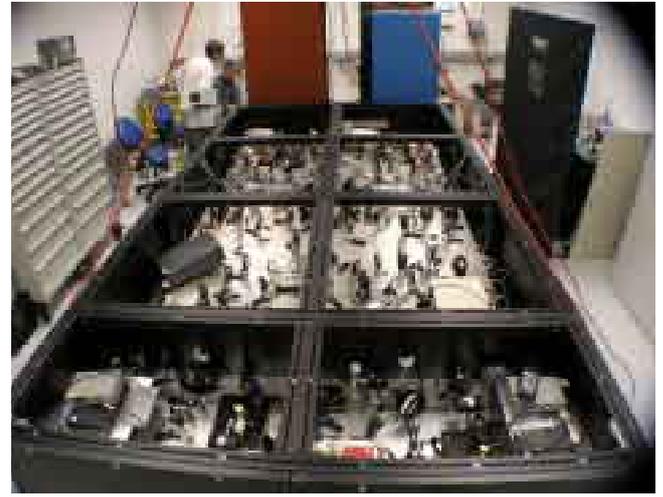

Figure 6.2: Two nulling interferometer testbeds. Left, the experiment by Ollivier et al. (2001) at IAS, Orsay. The superimposed red lines indicate the optical path. An input beam is divided into 2 parts to simulate the light coming from 2 different spacecraft. The beams recombine in destructive or nulling interference mode. Right, The Planet Detection Testbed at JPL, which simulates a bright star and a faint planet. The planetary signal can be extracted from the global flux when the contrast ratio is below 2 million.

goals will be met with continued effort and investment.

Additional key technological developments in recent years include:

- Selection of the baseline *interferometer configuration*. Significant effort in this area since 2000, backed by independent studies in Europe and the US, has identified the non-planar Emma X-Array as the optimal choice (see section 4.2);

- *Achromatic Phase Shifters (APS)*, which allow broadband destructive interference between beams, have reached TRL 4. A comparative study currently running in Europe should identify the preferred approach;

- *Space-qualified Delay Lines* to balance the different optical paths to nanometre accuracy have been demonstrated to TRL 5. A breadboard at TNO-TPD has achieved this performance at 40K and may be included as a test payload in the PROBA 3 space mission (section 6.2);

- *Single Mode Fibres*, or *Integrated Optics Modal Filters* that enable broadband nulling are now at TRL 4. Chalcogenide fibres have demonstrated the required performance of 40% throughput and 30 dB rejection of higher order spatial modes. Ongoing work is emphasizing silver halide single-mode filters, which will operate in the 12-20 μm band. Photonic Crystal fibres, which can cover the whole spectral domain in a single optical channel are being considered;

- *Detector Arrays* with appropriate read noise and dark current are at TRL 5/6. The Si:As Impurity Band Conductor (IBC) arrays developed for JWST appear to be fully compliant with *Darwin* requirements. A reduced-size version of the JWST 1024 x 1024 detector, e.g. 512 x 8 (300 x 300 for the general astrophysics program), could be read out at the required rate with a dissipation of a few tens to hundreds of μW. These devices exhibit high quantum efficiency (80%), low read noise (19 $e^-$), and minimal dark current (0.03 $e^-$/s at 6.7 K). Such performance permits sensitive observations, even at moderately high spectral resolution (*Res* = 300);

- Low vibration *Cryo-coolers* for the detector system are now at TRL 4. A European TRP has led to a prototype sorption cooler providing 5 mW of cooling power at 4.5 K. JPL scientists have demonstrated a system with 30 mW of cooling at 6 K.

*The message from the last decade of Darwin technology development is clear: if the Research and Technology effort that has been pursued in both Europe and the United States continues vigorously, Darwin's technology will reach TRL 5/6 by 2010, allowing it to be selected as ESA's first L mission for launch in 2018-2020.*

## 6.2 Precursor Missions

### 6.2.1 Exoplanet Discovery and Statistics

*COROT* (in operation)

COROT is a CNES led mission that is searching for planetary transits. It was launched at the end of 2006 and commissioning is running very successfully. For example, an excellent quality transit light curve has already been recorded. COROT has a 27 cm off-axis telescope and will observe ≥ 5 fields with about 12,000 stars each for a period of 5 months. COROT can detect planets with $R_{pl} \geq 2\ R_\oplus$ and orbital periods $\leq 50$ days. As early as 2008-2009, COROT should provide statistics on these objects and, by extrapolation, information on the abundance of terrestrial planets in the Habitable Zone.

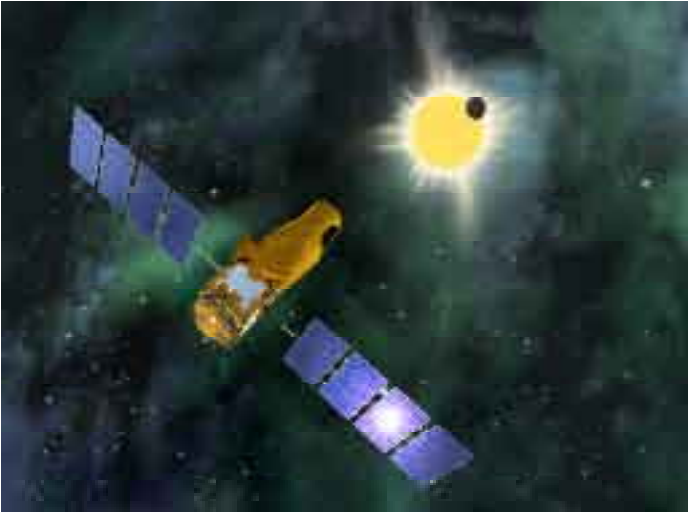

Figure 6.3: The COROT mission can detect hot big earths

*Kepler* (under construction)

Scheduled for 2008-2009, *Kepler* will detect terrestrial and larger planets near the HZ of stars with a wide variety of spectral types (Koch et al., Bull. of the AAS, 30, 1058, 1998). Its 0.95 m diameter telescope, pointed continuously at a single field, will monitor about 100,000 main-sequence stars located a few hundred parsecs away with the precision to detect Earth-sized planet transits. Over its 4 year lifetime, *Kepler* should provide the statistical abundance of the type of terrestrial planets that *Darwin* aims to characterize. This information will be valuable for *Darwin* mission planning.

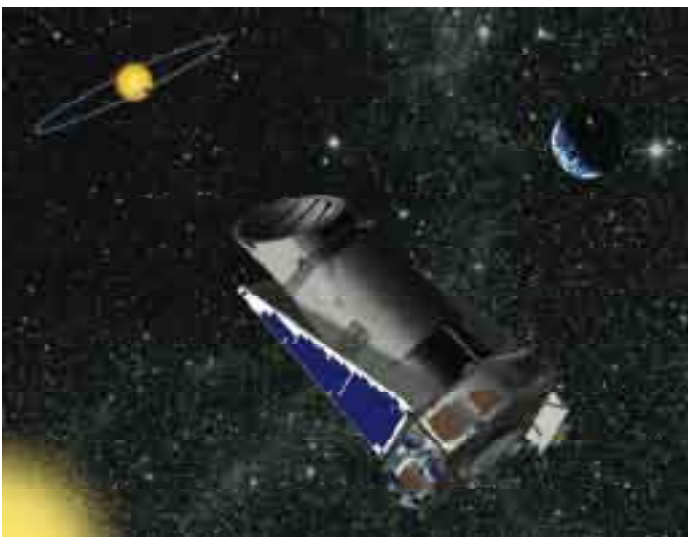

Figure 6.4: Kepler can detect Earth-like planets.

### 6.2.2 Formation Flying

The *Darwin* interferometer relies on Formation Flying (FF) technology to control the four collector spacecraft and one beam combiner. Section 4 describes how this strategy offers significant advantages. As with any new approach, however, FF should be validated in space. Europe has initiated several precursor missions:

*PRISMA* (approved)

PRISMA is a Swedish-led technology mission, which intends to demonstrate FF and rendezvous technologies, bringing them to TRL 8/9 (Figure 6.5). The Swedish Space Corporation is leading this ef-fort, which is funded by Sweden, Germany, Denmark, France and Alcatel. PRISMA comprises two spacecraft and should be launched in autumn 2008 into a low, Sun-synchronous orbit (600-1000 km) with a mission lifetime of about 8 months. The main objectives are to carry out technological flight demonstrations and manoeuvring experiments, including guidance, navigation, control, and sensor techniques (Persson and Jacobsson, IAC-06-D1.2.02, 2006). The positioning of the spacecraft relies on an Alcatel relative GPS technology, which should have an accuracy of ~10 cm. For inter-satellite distances less than 6 m, additional optical metrology should improve this accuracy.

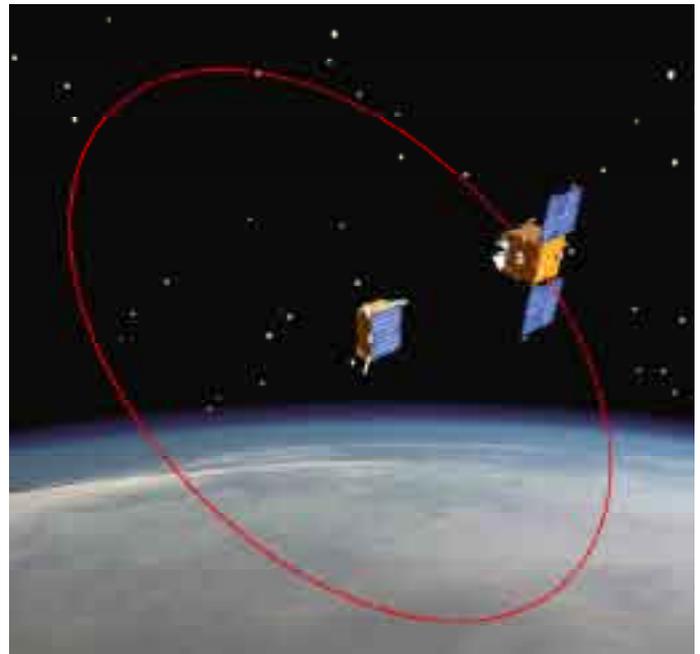

Figure 6.5: Artist's rendition of PRISMA. The larger of the two spacecraft carries most of the equipment and orbits around the smaller vehicle.

*PROBA 3*

The PROBA-3 mission is the next logical step, building on PRISMA's achievements. In addition to RF metrology, PROBA-3 will demonstrate 30 μm relative positioning accuracy, bringing optical metrology sensors to TRL 8/9. PROBA 3 could be launched in 2010, but it is not yet fully funded. This mission is not essential for *Darwin*, but we favour its completion, because it would provide further in-space demonstration of FF technologies.

*Pegase/ PERSEE*

*Pegase* is a single Bracewell interferometer that was proposed in the framework of the 2004 call by CNES for its formation flying demonstrator mission. The main scientific goal of *Pegase* is the high-angular resolution study of extrasolar giant planets at near-infrared wavelengths (2.5 to 5 μm). The mission could be extended to the study of brown dwarfs, circumstellar disks and dust tori around active galactic nuclei (AGN). CNES performed a Phase 0 study, but the mission was not selected for budgetary reasons. The construction of a mission breadboard, called PERSEE, is under consideration to increase our understanding of nulling interferometers. Note that, in order to ensure knowledge transfer to ESA, *Pegase* is being proposed as an M mission within the Cosmic Vision programme. *An attractive possibility would be a merger with PROBA 3, with additional European and possibly international participation. This would allow the inclusion of stellar interferometry into the mission.*

## 7. Cost plan and International Cooperation

### 7.1 Cost estimate

*Darwin* is proposed as an L mission. *Darwin* will be one of the most ambitious missions that ESA has ever undertaken, and we emphasize that *a phase A study is needed to derive an accurate cost estimate*. Here, we present a preliminary estimate, based on information from Alcatel (now Thales Alenia Space, TAS), Astrium, and a recent study of the Emma X-array by JPL.

The introduction of the Emma concept (Karlsson, Proc. SPIE 5491, 831, 2004; TAS, 2006) provides *a major simplification of the instrument*, eliminating all deployable components except the antennas. The optics of the Collector Spacecraft (CS) is reduced to a single mirror (section 4.2). Note that the CS cost scales as a function of the mirror diameter, because it is the main component of the spacecraft.

We present the costing for a 2 m version as a baseline, but also show the cost and performance trade for *Darwin* with mirrors of diameter 1.5, and 1.0 m. The corresponding numbers of stars screened and planets studied change, but at *constant quality per object* (section 4.3.3).

Further cost savings may be achievable in the design of the Beam Combiner Spacecraft (BCS). For instance, the current baseline design requires two distinct optical paths, because of the limited spectral range of existing single-mode fibres. However in the near future, Photonic Crystal fibres may allow operation over the entire 6-20 μm range with a single optical path.

*Baseline Cost Items for the Emma X-array configuration with 2 m collecting mirrors:*

- *Flight elements.* We use the JPL estimate of spacecraft masses plus a 20% margin, and apply a mean cost of 220 k€ per payload kg, which includes 15% contingency. These values are typical for scientific payloads (Earth and astronomical observations) as well as for telecommunication satellites, (information supplied by a European prime contractor). Total: 845 M€.
- *Launcher.* Ariane V ECA, 6.6 tons deliverable to L2: 125 M€ (CV Annex 4).
- *Ground Segment (5 years operations).* Considering the volume of communications, a prime contractor estimates the ground segment cost at 55 M€, which seems conservative when compared to that of GAIA (48 M€).
- *Pre-implementation and Space Agency internal costs.* We apply 1% and 11% of the total, respectively, as required in CV Annex 4.
- *Contingency.* This is already included: 20% on mass plus 15% on cost per kg

The total cost for 2 m collectors is then : 845 + 125 + 55 + 12 + 128 = 1165 M€ ≈ **1200 M€**.

Table 7.1 presents cost and performance estimates for different CS mirror sizes, using a conservative scaling dependence of the collecting mirror mass with their diameter, $M \propto D^2$.

### 7.2 International Cooperation

*Darwin* science has worldwide appeal. At present, both *NASA* and the Japan space agency, *JAXA*, have indicated their interest in the mission and their willingness to participate in the study phase (see the attached Letters of Commitment). Their letters also indicate a possible participation in the construction and operation phases. Contacts with other agencies are being cultivated.

With their parallel TPF-I studies over the last decade, NASA represents a powerful partner to ESA

Table 7.1: Estimated total cost and science performance for Darwin with different Collector Spacecraft mirror diameters. The first row corresponds to the baseline choice of 2.0 meter mirrors.

| Collector Diameter (m) | Mass (kg) | **Total cost (M€)** | Number of Screened Stars | Planets with $O_3$, $CO_2$ spectra | Planets with $H_2O$ spectra.[1] |
|---|---|---|---|---|---|
| 2.0 (baseline) | 3 830 | **1 200** | 218 | 49 | (24) |
| 1.5 | 2 960 | 950 | 142 | 32 | (18) |
| 1.0 | 2 290 | 800 | 76 | 17 | (14) |

[1] The method for detecting $H_2O$ is currently being re-assessed. Therefore, the number of planets for which this molecule can be searched is preliminary.

in the *Darwin* effort. Japan has considerable expertise in several key fields, including cryogenics with the AKARI mission, and mirror engineering.

## 7.3 European Networks

Interdisciplinary studies which focus on the origin and evolution of the atmospheres of terrestrial planets open a great opportunity for scientists with expertise in astrophysics, planetology, atmospheric physics and chemistry, climate physics, space plasma physics, magnetospheric physics, biophysics and biology.

The *Darwin* precursor science program will be extended to a large scientific community. It will also be performed via networking activities coordinated by discipline working groups within the EU *EUROPLANET* project [http://europlanet.cesr.fr/], exoplanets, and habitability related *ISSI* teams [http://www.issibern.ch/], the European Interferometry Initiative, the European Astrobiology Network Association (*EANA*), as well as nationally based research centres. The results of these studies will be presented to the wider scientific community via conference sessions and pursuing the *Darwin / TPF conference series*, e.g. the fifth scheduled in Naples, March 10-14, 2008.

## 8. Public Outreach

*Darwin* science transcends the narrow interest of typical scientific enquiry. When the mission succeeds in identifying another world like our own, simply everything will change, from science to politics to religion. Needless to say, this type of investigation has a profound appeal to the general public. As a result, *Darwin* has both the opportunity and responsibility to support a significant public outreach programme throughout all phases of the mission.

## 8.1 Pre-Launch Activities

*Darwin* scientists have already been involved in outreach activities, including:

- Interviews and articles in newspapers and magazines: several hundred in Europe, as well as Russia, Canada, and the USA
- Radio interviews: about 30 in Holland, Scandinavia and the UK
- Television programs and interviews: more than a dozen in a variety of countries. These include networks with worldwide reach, such as the BBC and the Discovery Channel.
- Public lectures: numerous presentations at universities, high schools and at the Hague Model United Nations

Additional plans include addressing school children and carefully designed exhibits in museums. For example, we foresee a touring exhibit with a theme of Life in the Universe. Ideally, such an endeavour would involve ESA as a leading and/or sponsoring partner.

In combination with these exhibits, contests could be arranged in which school classes would compete for the best original idea on how to exploit *Darwin*. The winning programme would then be implemented and executed.

Numerous opportunities exist to meet the public face-to-face and to make presentations in connection with scientific meetings. This is also already happening at the regular *Darwin* - TPF conferences held alternatively in Europe and the United States. Inviting local celebrities, for example well-known scientists or astronauts, can increase the impact of these activities.

## 8.2 Outreach During Science Operations

As with a typical space mission, ESA together with the science team would hold regular press conferences, issue press releases, etc. Note, however, that the "images" obtained with *Darwin*, whether in the nulling or imaging mode, will not have the visual appeal of Hubble Space Telescope imagery. This calls for particular competence and skill in clarifying the impact of science results as they come in. Presented properly, even a relatively poor image of another Earth can be transformational.

## 8.3 Post-mission activities

We anticipate collection of the major results and dissemination via the Internet, on DVDs, etc. *Darwin* is hopefully only the first of a long line of space missions that will investigate other worlds like our own. A carefully managed post-mission outreach program will ensure continued public support for exoplanet science.

# 9. List of Proposers

## 9.1 Core Proposers

<u>Principal Investigator</u>: Léger A. (F)    e-mail: Alain.Leger@ias.fr

<u>Astronomy</u>: Baglin A. (F) ; Beichman C. (USA) ; Coudé du F. V. (F) ; Danchi W. (USA); Eiroa C. (E) ; Greaves J. (UK) ; Henning T. (D) ; Jones H. (UK) ; Johnston K. (USA) ; Liseau R. (S) ; Malbet F. (F) ; Mennesson B. (USA) ; Mourard D. (F) ; Moutou C. (F); Nelson R. (UK); Paresce F.(I); Röttgering H. (NL) ; Rouan D. (F) ; Shibai H. (Japan);  Tamura M. (Japan) ; White G. (UK).

<u>Planetary Science</u>: Benz W. (CH) ; Blanc M. (F) ; Colangeli L. (I) ; Kaltenegger L. (A) ; Lammer H. (A), Ollivier M. (F); Paillet J. (F, NL); Quirrenbach A. (D) ; Stam D.(NL); Tinetti G. (F) ; Westall F. (F).

<u>Exobiology – Biosignatures, Chemistry</u> : Brack A (F) ; Cockell C. (UK) ; Cottin H. (F) ; d'Hendecourt L. (F) ; Schneider J. (F) ; Selsis F. (F) ; Westall F. (F).

<u>Instrumentation and Data Reduction Scientists</u>: Absil O.(B) ; Chazelas B. (F) ; Chelli A. (F); Defrère D. (B); den Herder J-W. (NL) ; Herbst T. M. (D) ;  Kern P. (F) ; Labadie L. (D)  Launhardt R. (D) ; Lawson P. (USA) ; Lay O. (USA); LeDuigou J-M. (F); Martin S. (USA) ; Mawet D. (B) ; Mourard D. (F) ; Mugnier L. (F) ; Queloz D. (CH) ; Rabbia Y. (F) ; Santos N. (P); Serabyn G. (USA) ; Thiébaut E. (F).

## 9.2 Supporting Proposers

The following 650 scientists support the *Darwin* proposal. After each name, their particular field of expertise is shown as follows: Astronomy (F 1), Comparative Planetology (F 2), Geophysics (F2.5), Biology-Biosignatures (F 3), Instrumentation and Data Reduction (F 4), Public Outreach (F5)

**Nobel Prize Winners**

| | | |
|---|---|---|
| Blumberg, B. (Medicine, 1976) | Perl, M. L. (Physics, 1995) | Townes, C. (Physics, 1964) |

**Australia**

Campbell J.(J. Cook Univ., Caims; F3)

**Austria**

| | | |
|---|---|---|
| Biernat, H. K., (SRI, Graz; F2) | Kerschbaum, F. (U-Vienna; F1) | Panchenko, M. (SRI, Graz; F2) |
| Dvorak, R., (U-Vienna; F2) | Lebzelter, T. (U-Vienna; F2) | Rucker, H. O. (SRI, Graz; F2) |
| Eichelberger, H. U(SRI, Graz; F1) | Leitzinger, M. (U-Graz; F1) | Stan-Lotter, H. (U-Salzburg; F3) |
| Fendrihan, S., (U-Salzburg; F3) | Lichtenegger, H., (SRI., Graz; F2) | Stangl, G., (SRI, Graz; F2) |
| Hanslmeier, A., (U-Graz; F1) | Lohinger, E., (U-Vienna; F2) | Weiss, W. W. (U- Vienna; F1) |
| Hron, J. (U- Vienna; F1) | Oberhummer, H., (TU-Vienna) | Zeilinger, W. W., (U- Vienna; F1) |
| Khodachenko, M. L., (SRI, Graz; F4) | Odert, P. (U-Graz; F1) | |

**Belarus**

| | | |
|---|---|---|
| Krot, A. M. (I. Informatics, Minsk; F1, F2) | Minervina, H. B(I. Informatics, Minsk; F1, F2) | Tkachova, P. P. (State University,Minsk F1, F2) |

**Belgium**

| | | |
|---|---|---|
| Borkowski V. (ULg; F1, F2) | Herman M. (ULBruxelles; F1, F2, F3) | Manfroid J. (ULg; F1) |
| Claeys, P. (Vrije Univ Brussel; F2); | Herwats E. (ULg; F1, F4) | Muller C. (I. Space Aeoronomy; F2) |
| Defise J-M. (CSL; F4) | Javaux, E. (Univ. de Liège; F3) | Namsaraev Z. (ULg; F3) |
| Dehant, V. (Royal Observ.; F2); | Jehin E. (ULg; F1, F2) | Sterken, C. (Vrije U. Brussel; F1, F2); |
| Galleni M. (ULg; F3) | Karatekin O. (ROB; F2) | Surdej, J. (Univ. de Liège; F1); |
| Hanot, C. (ULg; F4) | Leys N. (BNC; F3) | Vanaverbeke S. (KU Leuven; F1) |
| | | Verhoelst T. (KU Leuven; F1, F3) |

**Canada**

McCutcheon B. (UBC; F1)

**Chile**

| | | |
|---|---|---|
| Le Bouquin J-B. (ESO, Santiago; F1, F3) | Morel S. (ESO, Santiago; F3) | Schoeller M. (ESO, Santiago; F3) |

## China

Liu J. (NAOC, Beijing; F2, F3)  
Ouyang Z. (CCEP for Moon; F2; F3)  
Sun K. (U-Hong-Kong; F1)  
Zhuang, F. Y. (Biosc. and Bioengin., NAOC, Beijing; F3

## Croatia

Pecnic, B., (Dept. of Physics, U-Split)

## Denmark

Andersen A. (U-Copenhagen)  
Hinse T.C. (U-Copenhagen; F1)

## Finland

Kallio, E., (Meteo., Helsinki; F2)  
Hanski M. (Tuorla Obs.)  
Lehto K. (Tuorla Obs.)  
Mattila K. (U-Helsinki)

## France

Abe L. (OCA, Nice; F4)  
Aime C. (U-Nice; F4)  
Allard F. (ENS, Lyon; F1)  
Allemand P. (ENS, Lyon; F2)  
Aristidi E. (U-Nice; F1, F4)  
Augereau J-C. (U-Grenoble; F1)  
Barge P. (LAM, Marseille; F1)  
Benest D. (OCA, Nice; F1)  
Bénilan Y. (U-Paris; F2, F3)  
Berger J-P (LAOG; F1, F4)  
Bibring J-P (IAS, Orsay; F2)  
Blazit A. (OCA, Nice; F1, F4)  
Bonneau D. (OCA, Nice; F1, F4)  
Bordé P. (IAS, Orsay; F1, F4)  
Bourdieu L. (ENS biologie,Paris; F3)  
Brunier S. (Journalist, Sc. & Vie; F5)  
Cabane M. (Aero.,Paris; F2)  
Carbillet M. (OCA, Nice; F4)  
Casoli F. (IAS, Orsay; F1)  
Cassaing F. (ONERA, Paris; F4)  
Catala C. (Obs. Paris; F1, F4)  
Cavalié T. (LAB, Bordeaux; F2)  
Charnoz S. (CEA, Saclay; F1, F2)  
Chassefière E. (IPSL, Paris, F2)  
Chauvin G. (U-Grenoble; F1, F4)  
Chedid M. (LAOG; F1, F4)  
Chesneau O. (OCA, Nice; F1, F4)  
Cirou A. (chief ed. Ciel & Espace; F5)  
Courtin R. (Obs. Paris; F2)  
Coustenis, A. (Obs. Paris; F2)  
Crovisier J. (Obs. Paris; F2)  
de Laverny P. (OCA, Nice; F1)  
Deleuil M. (LAM, Marseille; F1)  
Delfosse X. (U-Grenoble; F1, F4)  
Despois D. (U-Bordeaux; F2, F3)  
Djouadi Z. (IAS, Orsay; F2)  
Dobrijevic M. (U-Bordeaux, F2)  
Domiciano A. (U-Nice; F1, F4)  
Doressoundiram A. (Obs. Paris; F2)  
Dougados C. (U-Grenoble; F1, F4)  
Dutrey A. (U-Bordeaux; F1)  
Duvert G. (U-Grenoble; F1, F4)  
Engrand E. (CSNSM, Orsay; F2, F3)  
Foellmi C. (LAOG; F1)  
Forget F. (U-Paris; F2, F2.5, F3)  
Forterre P. (U-Paris, F3)  
Forveille T. (U-Grenoble; F1, F4)  
Fouqué P. (U-Toulouse; F1)  
Foy R. (ENS, Lyon; F1, F4)  
Gavor P. (IAS, Orsay; F1, F4)  
Garcia-Lopez P. (U-Paris; F3)  
Gay J. (OCA, Nice; F1, F4)  
Giraud T. (U-Paris; F3)  
Grießmeier, J.-M. (Obs. Paris ; F2)  
Guillermo M. ((U-Grenoble; F4)  
Guillot T. (OCA, Nice; F2)  
Herpin F. (LAB, Bordeaux; F4)  
Jasniewicz G. (U-Montpellier; F1)  
Kamoun P. (UNSA, Nice; F2)  
Kern P. (LAOG; F4)  
Kervella P. (LAOG; F1, F4)  
Labèque A. (IAS, Orsay; F3, F4)  
Langlais, B. (U-Nantes ; F2)  
Le Coarer E. (U-Grenoble; F4)  
Le Coroller H. (O.H.Provence; F4)  
Leblanc F. (IPSL, Paris; F2)  
Leblanc, F. (Aéro., Paris ; F2)  
Lebonnois S. (IPSL, paris; F2)  
Lecavelier A. (IAP, Paris; F1)  
Léger J-F.(ENS biologie,Paris; F3)  
Léna P. (Acad Sci, Paris; F4)  
Lestrade J-F. (Obs. Paris; F1, F2)  
Lognonné P. (IPG, Paris, F2.5)  
Lopez B. (OCA, Nice; F1, F4)  
Martin G. (U-Grenoble; F4)  
Mège D. (U-Nantes; F2)  
Michel P. (OCA, Nice; F2)  
Mignard F. (OCA, Nice; F1, F2)  
Monin J-L. (U-Grenoble; F1, F4)  
Montmerle T. (U-Grenoble; F1)  
Morbidelli, A. (OCA; F1, F2)  
Mousis O. (Obs. Besancon; F2)  
Namouni F. (OCA, Nice; F2)  
Perraut K. (U-Grenoble; F1, F4)  
Perrier C. (U-Grenoble; F1, F4)  
Perrin G. (Obs. Paris; F1, F4)  
Petrov R. (U-Nice; F1, F4)  
Plez B. (U-Montpellier; F1)  
Poulet F. (IAS, Orsay; F2)  
Robbe-Dubois S. (U-Nice; F4)  
Robutel, P. (I. Méca. Cél. ; F1, F2)  
Roques F. (Obs. Paris; F1, F2)  
Rousselot P. (Obs. Besançon; F2)  
Rousset G. (Obs. Paris; F4)  
Schmidt B. (U-Grenoble; F2)  
Sotin C. (U-Nantes; F2)  
Stee P. (OCA, Nice; F1, F4)  
Tallon M. (ENS, Lyon; F4)  
Tanga P. (OCA, Nice; F2)  
Thébault P. (Obs. Paris; F1)  
Thevenin F. (OCA, Nice; F1)  
Toublanc D. (U-Toulouse; F2)  
Vakili F. (U-Nice; F1, F4)  
Vauclair S. (LAT, Toulouse; F1)  
Wakelam V. (U-Bordeaux; F1)

## Germany

Afonso C. (MPI-A; F1)  
Blum J. (TU-Braunschweig; F1)  
Brandner W. (MPI-A; F1)  
Bredehöft J. (IAPC, U-Bremen; F3)  
Breuer, D. (DLR, Berlin; F2)  
Carmona A. (MPI-A; F1)  
Delplancke F. (ESO, Garching, F1)  
Dreizler S. (U-Goettingen; F1)  
Duschl W.( U-Kiel; F1)  
Elias N. (Obs, Heidelberg; F4)  
Erikson A. (DLR, Berlin; F2)  
Feldt M. (MPI-A; F1)

Franck S. (Clim. I., Potsdam; F2.5)
Gail H.-P. (U-Heidelberg; F1)
Gawryszczak A. (MPI-A; F1)
Geisler R. (Obs, Heidelberg; F1)
Glindemann A. (ESO, Garch., F1)
Graser U. (MPI-A; F1)
Grenfell, J. L., (DLR, Berlin; F2, F3)
Gruen E. (MPI Kernphys.; F1, F2)
Guenther, E. (Tautenburg; F1, F2)
Hatzes, A. (Tautenburg; F1, F2)
Hauschildt P. (U-Hamburg; F1)
Klahr H. (MPI-A, Heidelberg; F1)
Klessen R. (U-Heidelberg; F1)
Kley W. (U-Tubingen; F1)
Kornet K. (MPI-A; F1)
Krabbe A. (Astron., U-Cologne, F1)
Kraus S. (MPIfR, Bonn; F1, F4)
Krivov A. (Astrophys. I., Jena; F1)
Kuerster M. (MPI-A; F1)
Millour F. (MPIfR, Bonn; F1, F4)
Motschmann U.(TU-Braunsch.; F2)
Mundt R. (MPI-A, Heidelberg; F1)
Neuhaeuser R. (Jena; F1)
Neukum G. (U-Berlin; F2.5)
Paetzold M. (U-Cologne, F2)
Rauer, H. (DLR, Berlin; F2, F3)
Renner S. (DLR, Berlin; F2)
Rettberg P. (I. Aero. Medec.; F3)
Schegerer A. (MPI-A; F1)
Setiawan J. (MPI-A; F1)
Spurzem R. (U-Heidelberg; F1)
Staude J. (Ed. "Sterne & Weltraum";F5)
Stecklum, B. (Tautenburg; F1, F2)
Stenzel, O. (MPI, Katlenburg; F2)
Strassmeier K.G. (Potsdam; F1)
Tornow C. (DLR, Berlin; F2)
Tubbs R. (MPI-A; F4)
v.d. Ancker M. (ESO; F1, F4)
von Bloh W. (Clim., Potsdam; F2.5)
von der Luhe O. (Freiburg)
Wambsgans J. (U-Heidelberg; F1)
Weigelt G. (MPI-RA., Bonn; F1)
Wiedemann G. (U-Hamburg; F1)
Wittkowski M. (ESO; F1, F4)
Wolf S. (MPI-A, Heidelberg; F1)
Wuchterl, G. (Tautenburg; F1, F2)
Wurm G. (U-Muenster; F2)

## Greece

Chatzitheodoridis E. (U-Athens, F3, F4)
Varvoglis H. (U-Thessaloniki; F1, F2)

## Hungary

Berces A. (Biophys., H. Acad. Sc, Budapest.; F3)
Erdi B., (E.Univ., Budapest; F1, F2)
Ronto G. (Biophys., H. Acad. Sc, Budapest.; F3)
Suli A. (Biophys., H. Acad. Sc, Budapest.; F3)

## India

S K Saha S. (IfA, Bangalore; F4)
Mayank V. (TIFR, Mumbai; F1, F4)

## Ireland

Butler R. (Nat. U., Galway)
Golden A. (Nat. U., Galway)
Hanlon L. (UC, Dublin)
Jordan B. (Dunsink Obs.)
McKenna-Lawlor J. (NU Irland, Kildare; F2)
Pollaco D. (QUB, Belfast; F1)
Ray T. (Adv.St, Dublin)
Sanders I. (Geology, Dublin)
Shearer A. (Nat. U., Galway)
Smith N. (Cork I. Techn.)

## Israel

Mazeh T. (U-Tel Aviv; F1)

## Italy

Alcala J-M. (Capodim., Napoli; F1)
Biazzo K. (Obs. di Catania; F1)
Blanco C. (U-catania; F2)
Bonomo A. (Obs. di Catania; F1)
Brucato J.R. (Capod., Napoli; F3)
Busà I. (Obs. Catania; F1)
Campanella G. (U-Roma; F1)
Cecchi-Pestellini C.(O.Cagliari; F1, F2)
Chiraravella A. (U-Palerme; F3)
Covino E. (Capod., Napoli; F1)
Cutispoto G. (Obs. Catania; F1)
Dilani A. (U-Padua; F1, F2)
Esposito F. (Capod, Napoli; F2, F4)
Gai M. (INAF, Torino; F4)
Lanza A.F. (Obs. Catania; F1)
Ligori S. (INAF, Torino; F1, F4)
Messina S. (Obs. Catania; F1)
Mura A. (IISP, Rome; F2)
Ortolani S. (Astr. U-Padova; F1)
Pagano I. (Obs. Catania; F1)
Penz T. (O. Palermo; F2)
Pecorella W. (Moon Base IWG, F4)
Silvotti R. (Capodim., Napoli; F1)
Strazzulla G. (Obs. Catania; F3)

## Japan

Abe L. (NAOJ; F1)
Baba N. (U-Hokkaido; F1)
Ebizuka N. (U-Konan; F4)
Enya K. (JAXA/ISAS; F1)
Ida S. (U-Tokyo; F2)
Inutsuka S-I (U-Kyoto; F1, F4)
Kaifu N. (U. of Air, Jap. Sc. Council; F1)
Kataza H. (JAXA/ISAS; F1)
Kitamura Y. (JAXA/ISAS; F1)
Kobayashi K. (N.U. of Yokohama; F3)
Kokubo E. (NAOJ; F2)
Kurokawa T. (U-Tokyo; F4)
Momose M. (U-Ibaraki; F1)
Morino J. (NAOJ; F1)
Murakami N. (NAOJ; F4)
Nakagawa T. (JAXA/ISAS; F1)
Nakamura R. (AIST; F2)
Nishikawa J. (NAOJ; F4)
Sato B-E (Tokyo Inst. Techn.; F1)
Sugitani K. (Nagoya Univ. Coll.; F1)
Sumi T. (U-Nagoya; F1)
Takeda M. (U of telecom.; F4)
Tavrov A. (NAOJ; F4)
Terada, N. (NIICT, Tokyo; F2)
Ueno M. (U-Tokyo; F1)
Yamamoto T. (U-Hokkaido; F1)

## Mexico

Lachaume R. (UNAM, Morelia; F1)  
Segura A. (UNAM, Mexico; F1)

## Netherlands

Bottinelli S. (U-Leiden; F1, F3)  
Brandl B. (U-Leiden)  
de Jong T. (SRON; F1)  
de Vries C. (SRON; F1)  
Dominic C. (U-Amsterdam)  
Hermsen W. (SRON; F1)  
Hogerheijde M. (U-Leiden)  
Langevelde H-J. (U-Groningen; F1)  
Miley G. (U-Leiden)  
Morganti R. (U-Groningen; F1)  
Nijman W. (Geology; F2.5)  
Roelfsema P. (SRON; F1)  
Schwartz A. (Kun; F3)  
Snellen I. (U-Leiden)  
Spaans M. (U-Groningen; F1)  
Tinbergen J. (U-Groningen; F1)  
van der Hucht K. (SRON; F1)  
Van der Werf P. (U-Leiden)  
Waters R. (U-Amsterdam)

## Norway

Mollendal H. (U-Oslo)  
Nielsen C. (U-Oslo)  
Kaas A. (Nordic O.Tel.)

## Poland

Konacki M.  
Niedzielski A. (U-Torum; F1)  
Szuszkiewicz E. (U-Szczecin, F1, F2, F3)  
Wolszczan A.

## Portugal

Alves, J.(Astr. Hisp.-Alem, Almeria)  
Bonfils, X. (CfA, U-Lisbon)  
Correia A. (U-Aveiro)  
Garcia, P. (U-Porto)  
Lago, M.T. (U-Porto)  
Marques, P.V.S. (INESC, Porto)  
Moitinho A. (SIM, Lisboa)  
Rebordao, J. (INETI, Lisbon)  
Santos, F. (U-Lisbon; F1)  
Yun J. L. (U-Lisbon)

## Romania

Dobrota C. (U-Babes-Bolyai; F3)

## Russia

Belisheva, N. (Pol-Alp Botan. Garden I., Apatity; F3)  
Chugunov Y. (R. Acad. Sc., Novgorod; F1, F2)  
Demekhov, A. (R. Acad. Sc., Nizhny; F1, F2)  
Erkaev, N. (I. Comput. Mod., Krasnoyarsk; F2)  
Fomin, B. (U-St. Petersburg)  
Gordiets, B. (Lebedev PI, Moscow, F2)  
Kachanova, T. (U-St. Petersburg)  
Kislyakov, A. (Lobachevsky SU, Novgorod; F2, F3, F4)  
Ksanfomality, L (DPS, IKI, Moscow; F2, F3, F4)  
Kulikov, Y. (Polar GI, Murmansk; F2)  
Mareev, E. (IAP, Novgorod; F2)  
Mingalev I. (Polar GI, Apatity; F2)  
Mingalev O. (Polar GI, Apatity; F2)  
Mingalev V. (Polar GI, Apatity; F2)  
Mingaleva G. (Polar GI, Apatity; F2)  
Pavlov, A. (LNSP, St. Petersburg; F1, F2)  
Semenov, V. (SU, St. Petersburg; F1, F2)  
Shaposhnikov, V. (IAP, Novgorod; F1, F2)  
Shematovich, V. (IA, Moscow; F2)  
Zaitsev, V. (IAP, Novgorod; F1, F2)  
Zelenyi L. (IKI/Moscow; F2, F3, F4)

## Slovenia

Zwitter T. (U-Ljubljana; F1)

## Spain

Abia C. (U-Granada)  
Alberdi A. (I. Astrofis. Granada)  
Alonso R. (I. Astrofis. Tenerife)  
Alonso A. (C S I C, Madrid)  
Amado P. (I. Astrofis., Granada)  
Anglada Escude, G. (U-Barcelona)  
Aparicio A. (I. Astrofis., Tenerife)  
Baeza S. (U. del Pais Vasco)  
Balaguer L. (U-Barcelona)  
Barrado D. (INTA, Madrid)  
Bayo A. (INTA, Madrid)  
Bejar V. (Gran Tel., Tenerife)  
Belenguer T. (INTA, Madrid)  
Belmonte J-A. (I. Astrof., Tenerife)  
Blasco C. (INTA, Madrid)  
Bouy H. (IAC, Tenerife; F1)  
Cantó-Doménech (U-P.Valen.; F1)  
Caballero J-A. (MPI-A, Heidelberg  
Campo A. (Univers. de Alicante)  
Cardiel, N. (U. C, Madrid)  
Castillo A. (U. C, Madrid)  
Catalan S. (I. EE, Barcelona)  
Cenarro A. J. (U C, Madrid)  
Colina L. (CSIC, Madrid)  
Colome P. (IEE, Barcelona)  
Cornide M. (UC, Madrid)  
Crespo I. (UC, Madrid)  
Cueto, M. (INTA, Madrid)  
de Castro Rubio E. (UC, Madrid)  
Deeg H. (I. Astrofis., Tenerife)  
de Gregorio I. (ESO, Chile)  
Delgado A. (I. Astrofis., Granada)  
Diaz Catala E. (INTA, Madrid)  
Domingo A. (INTA, Madrid)  
Eibe M-T. (C A, INTA, Madrid)  
Eliche M. del C. (UC, Madrid)  
Fernandez M. (I. Astr., Granada)  
Fernandez Soto A. (U-Valencia)  
Fernandez-Fig. M.(UC, Madrid)  
Fuente A. (Obs. Astron. Nacional)  
Gallego Maestro J. (UC, Madrid)  
Cruz Galvez M. (UC, Madrid)  
Garcia, D. (IC, London)  
Garcia Bedregal, A. (CSIC, Madrid)  
Garcia Lopez, R. (I. Astr., Tenerife)

Garcia Miro C. (INTA, Madrid)
Garcia Torres, M. (INTA, Madrid)
Garcia Vargas M. (FSL, Madrid)
Garrido R. (I. Astrofis., Granada)
Garzon F. (I. Astrofis., Tenerife)
Gil A. (U. Complutense Madrid)
Gomez, J-F. (I. Astrofis., Granada)
Gorgas J. (U. Complutense Madrid)
Gonzalez B. (INTA, Madrid)
Guirado J-C. (U-Valencia)
Gutierrez Sanc. R. (INTA, Madrid)
Gutierrez Soto J. (U-Valencia)
Hirschmann A. (EEC, Barcelona)
Hoyos C. (U-Autonoma de Madrid)
Huelamo N. (INTA, Madrid)
Israelian G. (I. Astrofis., Tenerife)
Lara L. (I. Astrofis., Granada)
Licandro J. (I. Astrofis., Tenerife)
Lopez B. (U-Barcelona)

Lopez Moreno J. (I. Astr., Granada)
Lopez Santiago J. (UC, Madrid)
Maldonado J. (UC, Madrid)
Molina Cuberos G. (U-Murcia; F2)
Montañés-Rodríguez (Tenerife, F3)
Marcaide, J. (U-Valencia)
Martin Guerrero E. (I. A., Tenerife)
Martinez Arnaiz, R. (UC, Madrid)
Martinez Garcia, V. (U-Valencia)
Mass M. (INTA, Madrid)
Montes, D. (UC, Madrid)
Moll Moreno, V. E. (INTA. Madrid)
Montesinos, B. (CSIC, Madrid)
Morales Cald. M. (CSIC, Madrid)
Morales Duran C. (CSIC, Madrid)
Palle, P. (I. Astrofis., Tenerife)
Palau Puigvert A. (CSIC, Madrid)
Pascual S. (UC, Madrid)
Perez Gonzalez P. G. (UC, Madrid)

Rebolo R. (I. Astrofis., Tenerife)
Regulo C. (I. Astrofis., Tenerife)
Reina M. (INTA, Madrid)
Ribas I. (I. Est. Espac., Barcelona; F1)
Risquez D. (IEE, Barcelona)
Rodriguez E. (I. Astrofis., Granada)
Rodrigo Blanco C. (CSIC, Madrid)
Rojas, J-F. (U-Pais Vasco)
Sabau D. (INTA. Madrid)
Sanchez Lavega A. (U-Pais Vasco)
Sanz Forcada, J. (CSIC, Madrid)
Sarro, L. M. (UNED, Madrid)
Solano E. (CSIC, Madrid)
Torrelles J.M. (IEE., Barcelona)
Trigo-Rodrig. J. M. (IEE Barcelona)
Villo, I. (U. Politecnica de C.)
Vitores, A. G. (U. Polit., Madrid)
Zamorano, J. (UC, Madrid)
Zapatero, M. R. (I. Astr., Tenerife)

### Sweden

Aalto S. (Onsala Obs.)
Björneholm O. (U-Uppsala)
Black J. (Onsala Obs.)
Booth R. (Onsala Obs.)
Brandeker A. (Stockholm O.)
Brandenburg A. (Stockholm O.)
Fransson C. (Stockholm O.)
Gahm G. (Stockholm O.)
Gumaelius L. (Stockholm HS)
Gustafsson B. (Uppsala Obs.)
Hallbeck L. (U-Goteborg, F3)
Heiter U. (Uppsala Obs.)
Hjalmarson A. (Onsala Obs.)

Hode T. (SW Natural Hist.)
Höfner S. (Uppsala Obs.)
Holm N. (U-Stockholm; F2.5)
Horellou C. (Onsala Obs.)
Justtanont K. (Stockholm O.)
Karlsson E. (U-Uppsala)
Kuchukhov O. (Uppsala Obs.)
Liljenstrom H. (A. Biosyst.)
Lindqvist M. (Onsala Obs.)
Löfdahl M. (Stockholm O.)
Lundin, R. (IRF, Umea; F2)
Mizuno M. (Uppsala Obs.)
Näslund M. (Stockholm O.)

Nikolic' S. (Onsala Obs.)
Nordström B. (Lund Obs.)
Olofsson G. (Stockholm O.)
Olofsson H. (Stockholm O.)
Olsson S. (Agri. Uppsala; F3)
Orndahl E. (Uppsala Obs.)
Poole A. (U-Stockholm; F3)
Regandell S. (Uppsala Obs.)
Rydbeck G. (Onsala Obs.)
Sandqvist A. (Stockholm O.)
Westerlund B. (Uppsala Obs.)
Wramdemark S. (Lund Obs.)
Yamauchi, M. (IRF, Kiruna; F2)

### Switzerland

Alibert Y. (U-Bern; F2)
Benz A. (I. Astr., Zurich)
Jetzer P. (U-Zurich)

Lovis C. (Obs. Geneva; F1)
Mayor M. (Obs. Geneva; F1, F4)
Pont F. (Obs. Geneva; F1, F4)

Ségransan D. (Obs. Geneva; F4)
Udry S. (Obs. Geneva; F1, F4)
Wurz, P. (U-Bern; F2)

### Taiwan

Ip, W.-H. (I. Astr, N. Centr. Univ.; F2)

### Ukraine

Konovalenko, A. (Insti. of Radioastron., Kharkov; F1)

### United Kingdom

Aigrain S. (U-Exeter; F1)
Anderson D. (U-Keele; F1)
Aylward A. (UC, London; F2)
Bingham B. (U-Strathclyde; F1)
Binns C. (U-Leicester; F2)
Birch M. (UC Lancashire; F2)
Bowey J. (UC, London; F2)
Braithwaite N. (Open U.; F2)

Buckle J. (Cambrid. MRA; F1, F3)
Burchell M. (U-Kent; F2)
Burgdorf M. (JM, Liverpool; F4)
Burleigh M. (U-Leicester; F1)
Buscher D. (U-Cambridge; F4) Buseman H. (Open U.; F2)
Catling D. (U-Bristol; F2, F3)
Chiu K. (U-Exeter; F1)

Cioni M-R. (U-Hertfordshire; F1)
Cho J. (QM U-London; F2)
Clark S. (Open U.; F1)
Clarke C. (Cambridge IoA; F1)
Collier Cameron A.(S.Andrews; F4)
Crowther R. (U-Southamp.; F2, F4)
Davis, G. (JAC, Hilo; F1)
Debattista V. (U-C-Lancashire; F1)

Dominik M. (U-St Andrews; F1)
Driver S. (U-St Andrews; F1)
Dworetsky M. (UC, London; F1)
Fitzsimmons A. (U-Belfast; F2)
Franchi I. (Open U.; F2)
Glasse A. (ATC., Edinburgh; F1) Goodwin S. (U-Sheffield; F1)
Grady M. (Open U.; F2)
Graffagnino V. (RAL; F1)
Grande, M. (U-Wales, Aberystwyth; F2))
Greaves J. (U-St Andrews; F1)
Green S. (Open U.; F2)
Griffin M. (U-Cardiff; F2, F4)
Gronstal A. (Open U.; F3)
Haniff C. (U-Cambridge; F1, F4)
Harrison R. (RAL; F2, F4)
Haswell C. (Open U.; F1)
Hellier C. (U-Keele; F1)
Helling C. (U-St Andrews; F1)
Hills R. (Cambridge MRA; F1, F4)
Holdaway R. (RAL; F1)
Horne K. (U-St Andrews; F1)
Horner J. (Open U.; F2)
Hough J. (U-Glasgow; F4)

Irwin P. (U-Oxford, F2)
Jenkins J. (U-Hertfordshire; F1)
Jheeta S. (Open U.; F3)
Jones B. (Open U.; F1)
Kolb U. (Open U.; F1)
Lewis S. (Open U.; F2)
Lim T. (RAL; F1, F4)
Marshall J. (Open U.; F1)
Mason N. (Open U.; F2)
McBride N. (Open U.; F2, F4)
Morgan G. (Open U.; F3, F4 )
Morris A. (Open U.; F2)
Norton A. (Open U.; F1)
Nutter D. (U-Cardiff; F1)
Parnell J. (U-Aberdeen; F3)
Parry I. (Cambridge IoA; F1, F4)
Penny A. (U-Edinburgh; F1)
Pickering J. (I.C. London; F2)
Pollacco D. (U-Belfast; F2,F4)
Raven J. (U-Dundee; F3)
Rawlings J. (UC, London; F1)
Rice K. (U-Edinburgh; F1)
Richer J. (Cambridge MRA; F1)
Rowan S. (U-Glasgow; F4)

Rymer H. (Open U.; F2.5)
Sarre P. (U-Nottingham; F1, F3)
Serjeant S. (Open U.; F1)
Sidher S. (RAL; F2)
Simpson F. (RO Edinburgh;F1)
Smith A. (UC, London; F4)
Smith M. (U-Kent; F1, F2)
Spreckley S. ((U-Birmingham; F1)
Stevens I. (U-Birmingham; F1)
Swinyard B. (RAL; F1, F4)
Tilanus R. (JAC, Hilo; F1)
Urquhart J. (U-Leeds; F1, F4)
van Loon J. (U-Keele; F1)
Viti S. (UC, London; F1)
v.Fay-Siebenburgen R.(Sheffield; F1)
Walker H. (RAL; F1)
Wall R. (Astrium UK; F4)
Ward-Thompson D. (U-Cardiff; F1)
Wheatley P. (U-Warwick; F1)
Widdowson M. (Open U.; F2.5)
Williams I. (QM, London; F1, F2)
Wright I. (Open U.; F2)
Wyatt M. (Cambridge IoA; F2)
Young P. (RAL; F1)

## United States of America

Adams F. (Ann Arbor, Michigan)
Akeson R. (Michelson SC; F1)
Allen R. (STSI, F1)
Atreya, S. (U-Michigan; F2, F3)
Barry R. (NASA GSFC; F1, F4)
Boss A. (Carnegie; F1, F2)
Carpenter K (NASA GSFC; F1, F4)
Colavita M. (JPL; F1)
Crisp D. (JPL; F1)
D'Angelo G. (NASA Ames; F1)
Debes J. (Carnegie; F1)
Dorland B. (USNO; F1, F4)
Dressler A (Carnegie Inst.; F1)
Falkowski P. (Nat. Acad. Sc., U-Rutgers; F2, F3)
Fomalont E. (NRAO; F1, F4)
Ford E. (CfA; F1)
Gaume R. (USNO; F1, F4)
Hinz P. (U-Arizona; F1, F4)
Hollis J. (NASA/GSFC; F1, F3)
Hyland D. (Texas A&M; F1)

Kane S. (U-Florida, F1)
Kuchner M (NASA GSFC; F1, F4)
Kulkarni S. (Caltech; F1, F4)
Lane B. (MIT; F1, F4)
Laughlin G. (UC Sa. Cruz; F1, F3)
Lin D. (UC Sa. Cruz; F1, F2)
Luhmann, J. (U-Berkely; F2)
Marchis F. (UC, Berkeley; F2)
Marcy G. (UC Berkeley; F1)
McAlister H. (Georgia St. U.; F1)
Meadows V. (Astrobiology I.; F2, F3)
Millan-Gabet R. (Caltech; F1, F4)
Monnier J. (U-Michigan; F1, F4)
Muterspaugh M. (UC Berkeley; F1)
Nimmo F. (UC Sa. Cruz; F2)
Noecker C. (Ball Aerosp.; F1, F4)
Norman C. (JHU/STSI; F1)
Oliversen N. (USNO)
Owen T. (U-Hawaii, F2, F3)
Quillen A. (U-Rochester; F1)
Rajagopal J (NOAO; F1, F4)

Raymond, S.(U-Col., Boulder; F1, F2)
Rinehart S (NASA GSFC ; F1, F4)
Sasselov D. (Harvard, Cambridge)
Scalo J. (U-Texas; F1, F3)
Seager S. (MIT; F1, F2, F3)
Shao M. (JPL; F1, F4)
Showman A. (U-Arizona; F1, F2)
Shuller P. (CfA / IAS; F1, F4)
ten Brummelaar (CHARA; F4)
Tian, F. (U-Col., Boulder; F2)
Traub W. (JPL; F1, F2, F3, F4)
Tyson N.D. (Am. Museum Nat. History; F1, F3, F5)
Van Belle G. (Caltech; F4)
Welch W.J. (UC Berkeley; F1, F4)
Wilner D. (CfA; F1)
Woolf N. (U-Arizona; F1, F4)
Wootten A. (NRAO; F1, F4)
Wray J. (U-Cornell; F1, F2, F4)
Yelle, R. (U-Arizona, Tuc.; F2)